\documentclass[%
 reprint,
 amsmath,amssymb,
 aps,
 prc
]{revtex4-1}

\usepackage{graphicx}
\usepackage{dcolumn}
\usepackage{bm}
\usepackage[caption=false]{subfig}
\usepackage{booktabs}
\usepackage{csquotes}
\usepackage{dsfont}
\usepackage{nicefrac} 
\usepackage{makecell}
\usepackage{color}
\usepackage[T1]{fontenc}

\newcommand{\bra}[1]{\langle#1|}
\newcommand{\ket}[1]{|#1\rangle}
\begin{document}

\preprint{APS/123-QED}

\title{Core-polarization effects and effective charges \\ in O and Ni isotopes from chiral interactions}

\author{ Francesco Raimondi$^{1}$ and Carlo Barbieri$^{1}$}
 
\affiliation{%
 $^1$Department of Physics, University of Surrey, Guildford GU2 7XH, United Kingdom 
}%

\date{\today}

\begin{abstract}
\begin{description}
\item[Background] 
  Most nuclear structure calculations, even for full configuration interaction approaches, are performed within truncated model spaces. These require consistent  transformations of the Hamiltonian and operators to account for the missing physics beyond the active space, so that several recent efforts have been devoted to find compatible derivations of the effective operators.  The effective charges employed in the shell-model calculations, and fitted to reproduce experimental data, can be seen as the phenomenological counterpart of such renormalization for electromagnetic operators.

\item[Purpose] 
A coherent mapping of \textit{ab initio} approaches into shell-model valence spaces requires a consistent derivation of effective electromagnetic operators. 
Here, we make a first step to lay the bases for their microscopic derivation in the context of the Self-Consistent Green's Function approach.

\item[Methods]  
We compute electric quadrupole (E2) effective charges from microscopic theory by coupling the single-nucleon propagators to core-polarization phonons, derived consistently from a realistic nuclear interaction.   The nuclear correlations  are included nonperturbatively according to the third order Algebraic Diagrammatic Construction (ADC(3)) and the Faddeev random-phase approximation (FRPA). The  polarization effects are included by evaluating the Feynman diagrams that couple the internal multi-nucleon configurations to the single-particle transitions induced by the electromagnetic operator.

\item[Results]   
The effective charges for E2 static moments and transitions are computed for selected isotopes in the Oxygen ($^{14}\text{O}$, $^{16}$O, $^{22}$O and $^{24}$O) and Nickel ($^{48}$Ni, $^{56}$Ni, $^{68}$Ni and $^{78}$Ni) chains. The values found are orbital dependent especially for the neutron effective charges, which show also a characteristic decreasing trend along each isotopic chain. In general, the values are compatible with the phenomenological ones commonly used for shell-model studies in the $0p\, 1s\, 0d$ and $1p\,0f\,0g_{\frac{9}{2}}$ valence spaces.  

\item[Conclusion] 
The phenomenological shell-model effective charges can be explained through  \textit{ab initio} approaches,  where the sole experimental input comes from the fitting of the realistic nuclear interaction. Effective electromagnetic operators can be derived which are tailored for different valence spaces and for specific numbers of active nucleons.
\end{description}
\end{abstract}

\pacs{Valid PACS appear here}
\maketitle


\section{\label{intro}Introduction}

The shell-model~\cite{RevModPhys.77.427} allows manageable many-body computations of nuclei by assuming a few active nuclei
in relatively small valence spaces near the Fermi surface. This is possible because the coupling with core nucleons is energetically disfavoured at low excitation energies, so it can normally be neglected. However, when electromagnetic observables such as static moments or transitions are calculated, effective charges for neutrons ($e_{\nu}$) and for protons ($e_{\pi}$)  become necessary to reproduce the experimental data. In this picture, the effective charges result from polarization effects in the nuclear core  and  from virtual excitations to higher orbits. These are either induced by the valence nucleons undergoing the electromagnetic transition or by direct coupling of the electromagnetic field to out-of-valence particles. Thus, effective charges account for interactions with the nucleons that would be otherwise neglected because of the inert-core approximation.
Bohr and Mottelson~\cite{bohr1998} provided a version of this interpretation in terms of the coupling between the single-particle degrees of freedom and the collective oscillations associated with the deformation of the nuclear density in the core.

The effective charge describes in this way the coupling of the electromagnetic operator to nucleons outside the valence space, which can be extrapolated to the electromagnetic transitions for other excited states and to nuclei in a neighbourhood of the given mass region.
Besides the fact that effective charges are model-dependent quantities, their general validity  throughout a given major shell is an approximation. Simple considerations related to the presence of shell effects and to the additivity of the field exerted by the valence particle on the nucleons in the core, suggest that effective charges a) must be different for moments or transitions involving states with different single-particle quantum numbers, and b) should display a significant isospin dependence 
(see Ch.~3 of Ref.~\cite{bohr1998}). In fact, neutron-rich isotopes far from stability have weaker core polarization effects, due to the fact that extended wave functions of loosely-bound neutrons have reduced interaction with the well-bound nucleons in the core. For instance, in light open-shell nuclei such as C or Ne isotopes, calculated effective charges for neutron-rich nuclei are closer to the bare charges ($e_\nu = 0$ and $e_\pi = 1$) than the ones in the valley of 
stability~\cite{Sagawa2004}.  
This is also confirmed by comparison with measured electromagnetic moments for Boron and Lithium 
isotopes~\cite{PhysRevLett.69.2058}. 
Moving to heavier nuclei, the standard values for effective charges in $sd$-shell nuclei, $e_\pi \simeq 1.4$ and $e_\nu \simeq 0.5$, have been brought into question for neutron-rich isotopes belonging to the Aluminum, Carbon, and Oxygen chains. Smaller values than the standard ones have been 
proposed~\cite{Yuan2012, PhysRevC.79.054303,PhysRevC.79.011302, DeRydt2009344}.
Also for several isotopes calculated in the $pf$ valence space, the typical   values  of the effective charges $e_\pi \simeq $ 1.5 and  $e_\nu \simeq $ 0.5 
are not compatible  with values extracted from experimental static moments and electromagnetic
 transitions~\cite{duRietz2004,Brown1974,Hoischen2011,Allmond2014,Loelius2016}.

 Effective charges have been  derived microscopically by computing matrix elements of electromagnetic observables,  within particle-vibration coupling models based on microscopic phenomenological 
 interactions~\cite{SAGAWA198484,Hamamoto1996,Sagawa2004}. 
 In \textit{ab initio} configuration interaction calculations, one can relate electromagnetic observables computed within the full space of configurations with results in a restricted space by using proper effective operators: in this way, the impact of excluded space correlations is evaluated by computing the one-body and  two-body parts of the electromagnetic operator in the restricted valence space.   Effective  operators  have been obtained in this spirit by transforming the bare ones via an Okubo-Lee-Suzuki  transformation within the no-core shell 
 model~\cite{PhysRevC.55.R573,PhysRevC.80.024315}, or via  in-medium similarity renormalization group methods~\cite{PhysRevC.96.034324}.
For effective Hamiltonians, progress has been made in recent years by performing calculations of closed-shell isotopes within very large model spaces and mapping the resulting nucleon-nucleon in medium interactions into a shell-model 
valence space~\cite{Jansen2014prlOxSM,Bogner2014prlOxSM}. 
This has allowed to obtain effective interactions in the valence space from first principles, thus to apply \textit{ab initio} theory to a large number of (open shell) isotopes.  An analogous strategy has also been exploited earlier on within the Self-Consistent Green's Function (SCGF) approach to compute isotopes in the $pfg_{\frac{9}{2}}$ valence 
space~\cite{Barbieri2009prl}. 
However, that application was limited only to effective interactions and these were computed at the lowest order in the Feynman expansion, while higher order diagrams---and even all-orders resummations---might be needed for accurate 
computations~\cite{Morten1995EffInts}. 

The purpose of this work is to set an initial step towards deriving effective electromagnetic charges for shell-model applications by using  SCGF theory.  We start by considering the bare one-body  quadrupole electromagnetic (E2) operator. 
We define effective charges from the E2 matrix element that includes polarization correlations, i.e. induced by the E2 operator itself, for very large model spaces, and we take its ratio with the E2 matrix element with respect to uncorrelated states in the valence space of interest. In this sense, the ratio rescales the coupling strength (charge) of the bare electromagnetic operator, and renormalizes the operator matrix element computed in the restricted model space.
%
%
 The coupling of the single nucleon to the correlations in the nuclear core is achieved by adding virtual excitations to a reference state. That means dressing the single-particle propagation with the intermediate states configurations encoded in the self-energy. To renormalize the charge consistently with the chosen valence space, we block those virtual excitations made by the particle and hole configurations belonging to the considered valence space.  

The basic SCGF equations  are reviewed in Sec.~\ref{form}, while the formalism for the computation of the effective charges is introduced  in Sec.~\ref{form2}.
In Sec.~\ref{EffCH_calc} we present results for E2 effective charges of Oxygen ($^{14}$O, $^{16}$O, $^{22}$O and $^{24}$O in Sec.~\ref{Oxygen}) and Nickel ($^{48}$Ni, $^{56}$Ni, $^{68}$Ni and $^{78}$Ni in Sec.~\ref{Nickel}), whose typical shell-model valence spaces are the $0p\, 1s\, 0d$ and $1p\,0f\,0g_{\frac{9}{2}}$, respectively. 
Conclusions are drawn in Sec.~\ref{concl}.

\section{\label{form}Self-consistent Green's function formalism}
In this section we recall the basic definitions and equations of the SCGF formalism. Extensive reviews of this approach can be found in 
Refs.~\cite{Barb2017LNP,Dickhoff2004,dickhoff2005}.

We start from a many-body Hamiltonian composed by the kinetic energy term $\hat{T}$ and the two- (2NF) and three-nucleon forces (3NF) $\hat{V}$ and $\hat{W}$, i.e.
\begin{eqnarray}
\label{H}
\hat{H}_{int} &=& \sum_{\alpha \beta} T_{\alpha \beta} \, a^\dagger_\alpha a_\beta 
+
\frac{1}{4} \sum_{\substack{\alpha\beta\\\gamma\delta}}V_{\alpha\beta,\gamma\delta}\, a_\alpha^\dagger a_\beta^\dagger a_{\delta} a_{\gamma} \nonumber \\
&&+ 
\frac{1}{36}\sum_{\substack{\alpha\beta\gamma \\ \mu\nu\eta}} W_{\alpha\beta\gamma,\mu\nu\eta}\,
a_\alpha^\dagger a_\beta^\dagger a_\gamma^\dagger  a_{\eta} a_{\nu} a_{\mu}  \, ,
\end{eqnarray}
where we  use Greek indexes to label the states of the complete orthonormal single-particle basis that defines the computational model space. The subscript `$int$' indicates that we subtract the kinetic energy for the center of mass and the corresponding corrections are intended as already included in the matrix elements for~$\hat{T}$ and~$\hat{V}$.

The diagonalization of the Hamiltonian in Eq.~(\ref{H}) gives the exact many-body wave function of the system:
\begin{equation}
\label{Schro}
\hat{H}_{int} \ \ket{\Psi^{A}_n} = E_n^{A} \ \ket{\Psi^{A}_n} \, .
\end{equation}

The Hamiltonian $\hat{H}_{int}$ drives the dynamics of each nucleon interacting with the nuclear medium. This is captured by  the one-body propagator,
\begin{equation}
\label{Green}
 g_{\alpha \beta}(t-t') = - \frac{i}{\hbar}     \bra{\Psi^{A}_0} \mathcal{T}\left[ 	a_{\alpha}(t)  	a_{\beta}^{\dagger}(t') \right] \ket{\Psi^{A}_0} \, ,
\end{equation}
 with the time evolution of the field operators $a_{\alpha}$ and  	$a_{\beta}$ given in the Heisenberg picture, and their order specified by the  time-ordering operator $\mathcal{T}$.

It is useful to introduce the spectral representation of Eq.~(\ref{Green}),
\begin{align}
 g_{\alpha \beta}(\omega) ~={}& 
 \sum_n  \frac{ 
        \left( {\cal X}^n_{\alpha}  \right)^*
         {\cal X}^n_{\beta} 
              }{\hbar \omega - (E^{A+1}_n - E^A_0)+ \textrm{i} \eta }  \nonumber\\
 +{}& \sum_k \frac{
      {\cal Y}^k_{\alpha}
        \left({\cal Y}^k_{\beta}\right)^*
             }{\hbar \omega - (E^A_0 - E^{A-1}_k) - \textrm{i} \eta } \; ,
\label{Green_spectr}
\end{align}
that is expressed in the frequency domain via the Fourier transform of the time representation, Eq.~(\ref{Green}). The spectroscopic amplitudes in Eq.~(\ref{Green_spectr}) are given by
\begin{align}
{\cal X}^n_{\alpha} \equiv \langle\Psi_n^{A+1}  |a_{\alpha}^\dagger| \Psi_0^A  \rangle  \, , 
\nonumber \\
{\cal Y}^k_{\alpha} \equiv\langle\Psi_k^{A-1}|a_{\alpha}|\Psi_0^A\rangle \, .
\label{tran_ampl}
\end{align}

The exact solution of Eq.~(\ref{Schro}) entails the same physical content as the Dyson equation for the propagator:
\begin{equation}
  \label{eq:Dy}
g_{\alpha\beta}(\omega)=g^{(0)}_{\alpha\beta}(\omega)+ \sum_{\gamma\delta} g^{(0)}_{\alpha\gamma}(\omega)\Sigma_{\gamma\delta}^{\star}(\omega) g_{\delta\beta}(\omega)  \; .
\end{equation}
 The unperturbed propagator in Eq.~(\ref{eq:Dy}) is intended as being expressed with respect to a reference state $\phi^{A}_0$, 
\begin{equation}
\label{Green_0}
 g^{(0)}_{\alpha \beta}(t-t') = - \frac{i}{\hbar}     \bra{\phi^A_0} \mathcal{T}\left[ 	a^I_{\alpha}(t)  	{a^I_{\beta}}^{\dagger}(t') \right] \ket{\phi^{A}_0} \, ,
\end{equation} 
where the superscript `I' indicates the evolution of operators in the interaction picture.
 Eq.~\eqref{Green_0} describes the propagation of a single nucleon with kinetic energy~$\hat{T}$ and affected by an auxiliary mean-field one-body potential~$\hat{U}$.  The expansion of the correlated propagator in terms of the residual inter-nucleon interactions $-\hat{U}+\hat{V}$ and $\hat{W}$ is based on Eq.~(\ref{Green_0}) as the reference propagator.
 
To make the problem computationally tractable for the present application, we use as reference the Optimized Reference State (OpRS) propagator. This is obtained by reducing the number of poles of the dressed propagator, according to the procedure detailed in 
Ref.~\cite{Barbieri2009Ni56}. 
This is achieved by mapping the  fully correlated propagator of Eq.~(\ref{Green}) to a propagator with  the same number of poles as the mean-field state, while relevant sum rules obtained from the full propagator are forced to be fulfilled. The obtained OpRS propagator has the following 
form~\cite{Barbieri2009Ni56,Rocco2018},
\begin{align}
g_{\alpha\beta}^{\rm{OpRS}}(\omega)=\sum_{n \not\in F}\frac{(\psi^n_\alpha)^\ast\psi^n_\beta}{\hbar\omega-\epsilon^{\rm{OpRS}}_n+i\eta}+\sum_{k \in F}\frac{\psi^k_\alpha(\psi^k_\beta)^\ast}{\hbar\omega-\epsilon^{\rm{OpRS}}_k-i\eta} \, ,
\label{g_OpRS}
\end{align}
where  $\epsilon^{\rm{OpRS}}_{n/k}$ and $\psi_{\alpha}^{n/k}$ are the set of the single-particle energies and amplitudes, respectively. They form an orthonormal basis which is separated in occupied ($\in F$) and unoccupied ($\notin F$) states.

The $\Sigma_{\gamma\delta}^{\star}(\omega)$ in Eq.~(\ref{eq:Dy}) is  the irreducible self-energy that gives the coupling of single nucleon states to the virtual excitations in the nuclear medium. In our calculations, the latter are constructed as many-particle and many-hole configurations formed from the OpRS orbits $\psi_{\alpha}^{n/k}$ and interacting through 2NFs and 3NFs. The self-energy can be decomposed in a static part $\Sigma_{\alpha\beta}^{\infty}$ (local in time) and a dynamic part $\widetilde{\Sigma}_{\alpha\beta}(\omega)$ (energy dependent):
\begin{align}
\label{irr_SE_decomp}
\Sigma_{\alpha\beta}^{\star}(\omega) ={}& \Sigma_{\alpha\beta}^{\infty} ~+~  \widetilde{\Sigma}_{\alpha\beta}(\omega) 
\nonumber \\
&  \Sigma_{\alpha\beta}^{\infty}  ~+~ \frac 1 4 
 \sum_{\substack{\gamma \delta  \sigma \\ \mu \nu \lambda}} \tilde{V}_{\alpha \sigma, \gamma \delta}  \,
 R^{(2p1h/2h1p)}_{ \gamma \delta  \sigma, \mu \nu \lambda}(\omega) \,  \tilde{V}_{\mu \nu, \beta \lambda}
\, ,
\end{align}
where the effective interaction $\tilde{V}$ incorporates both the 2N and 
3NFs~\cite{Carbone2013PRC3nf,Raimondi2018ADC3}.

The dynamical part of the self-energy $\widetilde{\Sigma}_{\alpha\beta}(\omega)$ can be written in the Lehmann's spectral representation, as
\begin{align}
\label{irr_SE_Lehmann}
\widetilde{\Sigma}_{\alpha\beta}(\omega) \! & = \! \sum_{p  p'} \textbf{M}_{\alpha p}^\dagger \Bigg[ \frac{1}{ \hbar \omega \mathds{1} - (\textbf{E}^{>}   +  \textbf{C}) + \textrm{i} \eta\mathds{1}} \Bigg]_{\substack{\! \! p  p'}} \! \! \textbf{M}_{p' \beta}  \nonumber \\
 & \! + \! \sum_{q q'} \textbf{N}_{\alpha q} \Bigg[ \frac{1}{\hbar \omega \mathds{1} - (\textbf{E}^{<} +  \textbf{D}) - \textrm{i} \eta \mathds{1}} \Bigg]_{\! \!q q'} \! \! \textbf{N}_{q' \beta }^\dagger \, ,
\end{align}
with $p$ ($q$) being collective indexes for multiparticle-multihole intermediate state configurations (ISCs), beyond the single-particle (-hole) propagation.

\begin{figure}
    \subfloat{\includegraphics[scale=0.3]{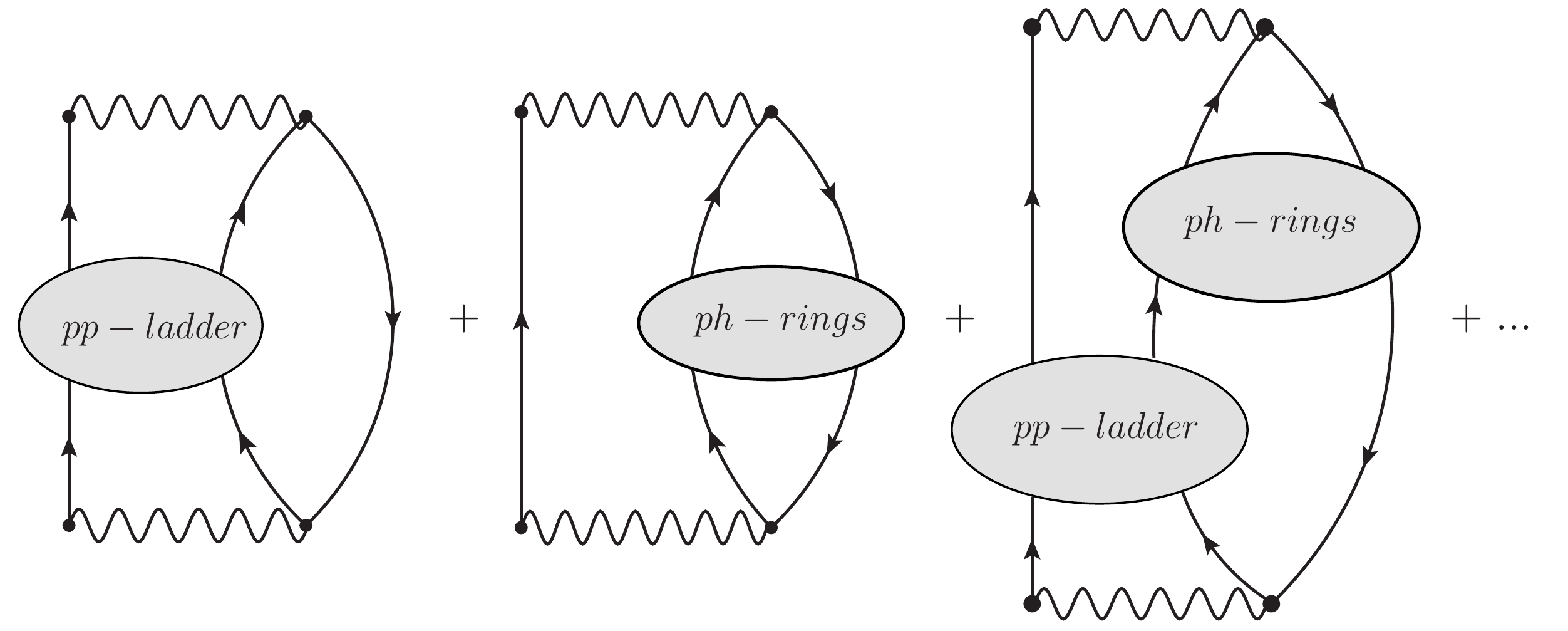}}
  \caption{
  Example of diagrams that contribute to the dynamic self-energy $\widetilde{\Sigma}_{\alpha\beta}(\omega)$ in Eq.~\eqref{irr_SE_decomp}. The wavy lines are two-body interactions $\widetilde{V}$ that incorporate both the 2NF and the normal ordered  3NF. Each ellipse represents an all-orders resummation of $pp$/$hh$ ladders of $ph$ rings.  The three-lines irreducible propagator $R^{(2p1h/2h1p)}(\omega)$ results from combining these phonons with single-particle states to all orders, in a Faddeev-like series.  }
  \label{FRPA_exp}
\end{figure}

In Eq.~(\ref{irr_SE_Lehmann})  we limit the forward-in-time indexes $p$ and $p'$  to two-particle-one-hole  ($2p1h$) ISCs. Analogously, we limit the backward-in-time indexes   $q$ and $q'$ to two-hole-one-particle  ($2h1p$) ISCs. 
Hence, only the three-lines irreducible propagator $R(\omega)$ enters Eq.~\eqref{irr_SE_decomp}.
The matrices $\textbf{M}_{p \alpha} $  and $\textbf{N}_{\alpha q} $ appearing in Eq.~(\ref{irr_SE_Lehmann}) are the coupling terms linking initial and final single-particle states to the propagation of $2p1h$  and $2h1p$ ISCs, respectively.
The $\textbf{E}^{>}$ and $\textbf{E}^{<} $ are the unperturbed energies of these ISCs and  $\textbf{C}_{\substack {p  p'}}  $  and $\textbf{D}_{\substack {q  q'}} $ are the interaction matrices among them. The denominators in Eq.~\eqref{irr_SE_Lehmann} imply all-order summations of phonons described by $pp$ and $hh$ ladders and $ph$ rings, as well as their interference and mixing to single-particle states. The diagrammatic content of $\widetilde{\Sigma}_{\alpha\beta}(\omega)$, and hence of $R^{(2p1h/2h1p)}(\omega)$, is sketched in Fig.~\ref{FRPA_exp}.

In the present work we evaluate the matrices entering Eq.~\eqref{irr_SE_Lehmann} in two different ways. In the standard third order Algebraic Diagrammatic Construction [ADC(3)] approach, the ladder and ring phonons are computed minimally by resumming diagrams in the Tamm-Dancoff Approximation 
(TDA)~\cite{RingSchuck}. The same exact resummation can be done all at once, by computing the inverse matrices in Eq.~\eqref{irr_SE_Lehmann}, or by pre-computing the single phonons and then recoupling them through a Faddeev series as depicted in Fig.~\ref{FRPA_exp}.  For this reason, the method is also referred to as the Faddeev-TDA (FTDA) approach.  The FTDA and ADC(3) approaches are exactly equivalent and the two names will be used interchangeably in the rest of this paper.
 It is also possible to compute the intermediate phonons in Fig.~\ref{FRPA_exp} using the Random Phase Approximation (RPA), which leads to the so called Faddeev-RPA (FRPA) 
 approach~\cite{Barbieri2001frpa,Barbieri2003ExO16,Barbieri2007Ne,Degroote2011mols}.
The ADC(3) is the method of choice for most modern computations of ground state properties and for studying the addition or removal of a nucleon because of its good accuracy and better numerical stability. However, the additional ground state correlations included into the RPA are known to have sizeable effects on the description of collective modes and they can be expected to impact the effective charges induced by these.  Both approaches involve non-perturbative all-orders calculations and we will compare their results in Sec.~\ref{EffCH_calc}.  The techniques to solve the Dyson equation as single eigenvalue problem, by the application of Lanczos-type algorithms, and other details of our implementation are discussed 
in Refs.~\cite{Soma2014,Barb2017LNP}.

\section{\label{form2} Effective charges from particle-vibration coupling}
On a fundamental level, effective charges  encapsulate the modification of the in-medium propagation of a nucleon in the presence of an external field.  In the most general case, the internal Hamiltonian $\hat{H}_{int}$ of Eq.~(\ref{H}) is then complemented with a time-dependent external field $\hat{\phi}(t)$,
according to
\begin{equation}
\hat{H}^{\phi}(t) = \hat{H}_{int} + \hat{\phi}(t) \, .
\label{H_phi}
\end{equation}
Typically, $\hat{\phi}(t)$ is  an electromagnetic field operator carrying angular momentum $\lambda$ and its projection $\mu_\lambda$,
 \begin{equation}
\hat{\phi}(t) \equiv e \; \hat{\phi}^{(\lambda \mu_\lambda)}(t) \, ,
\label{eq:t_field}
\end{equation}
where we have explicitly separated the electric charge $e$.

The nuclear many-body wave function acquires then a dependence on the external field, which is reflected in the Schr\"odinger equation,
\begin{equation}
\label{Schro_phi}
\hat{H}^{\phi} \ \ket{\Psi^{\phi  A}_n} = E_n^{\phi A} \ \ket{\Psi^{\phi A}_n} \, .
\end{equation}
The ground state solution of the many-body problem in Eq.~(\ref{Schro_phi}) will then allow to introduce the propagator in the presence of an external 
field~\cite{dickhoff2005},
\begin{equation}
\label{Green_phi}
 g^{\phi}_{\alpha \beta}(t-t') = - \frac{i}{\hbar}     \bra{\Psi^{\phi A}_0} \mathcal{T}\left[ a_{\alpha}(t)  a_{\beta}^{\dagger}(t') \right ] \ket{\Psi^{\phi A}_0} \, ,
\end{equation}
where in this case the Heisenberg picture is generated by $\hat{H}^{\phi}$.  One would then use time-dependent perturbation theory in $\hat{\phi}(t)$ to access the transition probabilities induced by the external field.

In this work, we rather pursue an heuristic approach  based on the physical insights given by the particle-vibration coupling 
model~\cite{bohr1998}. 
Hence, we describe the dynamics of the single nucleon as influenced by the collective vibrations of the surrounding nuclear medium, which in turn are induced by the external field. 
 Our goal will be to embed such dynamical effects into a single-particle space small enough that can be used in shell-model applications.
The first step is to define the shell-model valence space ($S_V$) we are targeting.  To this purpose we adopt the relevant OpRS orbits from the reference propagator~\eqref{g_OpRS} that results from the SCGF computation. Specifically the single-particle energies, $ \epsilon^i$, and wave functions, $|z^i\rangle = \sum_\alpha |\alpha\rangle z^i_\alpha$, are~%
\begin{align}
 z^i_{\alpha} ={}&
 \begin{cases}
     (\psi^n_\alpha)^\ast \\ 
     \; \psi^k_\alpha 
  \end{cases}
\hbox{~   and  \quad}
  \epsilon^i =
\begin{cases}
  \epsilon^{\rm{OpRS}}_n \quad &\hbox{\quad if $i$=$n \not\in F$  ,  } \\
  \epsilon^{\rm{OpRS}}_k &\hbox{\quad if $i$=$k \in F$  ,}
  \end{cases}
 \label{eq:OpRS_MSP}
\end{align}
where each orbit could be either occupied or unoccupied in Eq.~\eqref{g_OpRS} and we introduced a generic index $i$ for referring to both cases. 
Note that, in the same spirit of 
Refs.~\cite{Barbieri2009prl,Stroberg2017}, 
the nucleus $|\Psi_0^A\rangle$ to which the propagator~\eqref{Green} is associated does not need to be identified as the core nucleons of the shell-model space. In fact, one can assume a given valence space $S_V$ and target a specific nucleus inside it. The  effective charges extracted from our calculations would then be tuned to that particular region of the nuclear chart.

\begin{figure}[t]
  \centering
    \subfloat{\includegraphics[scale=0.50]{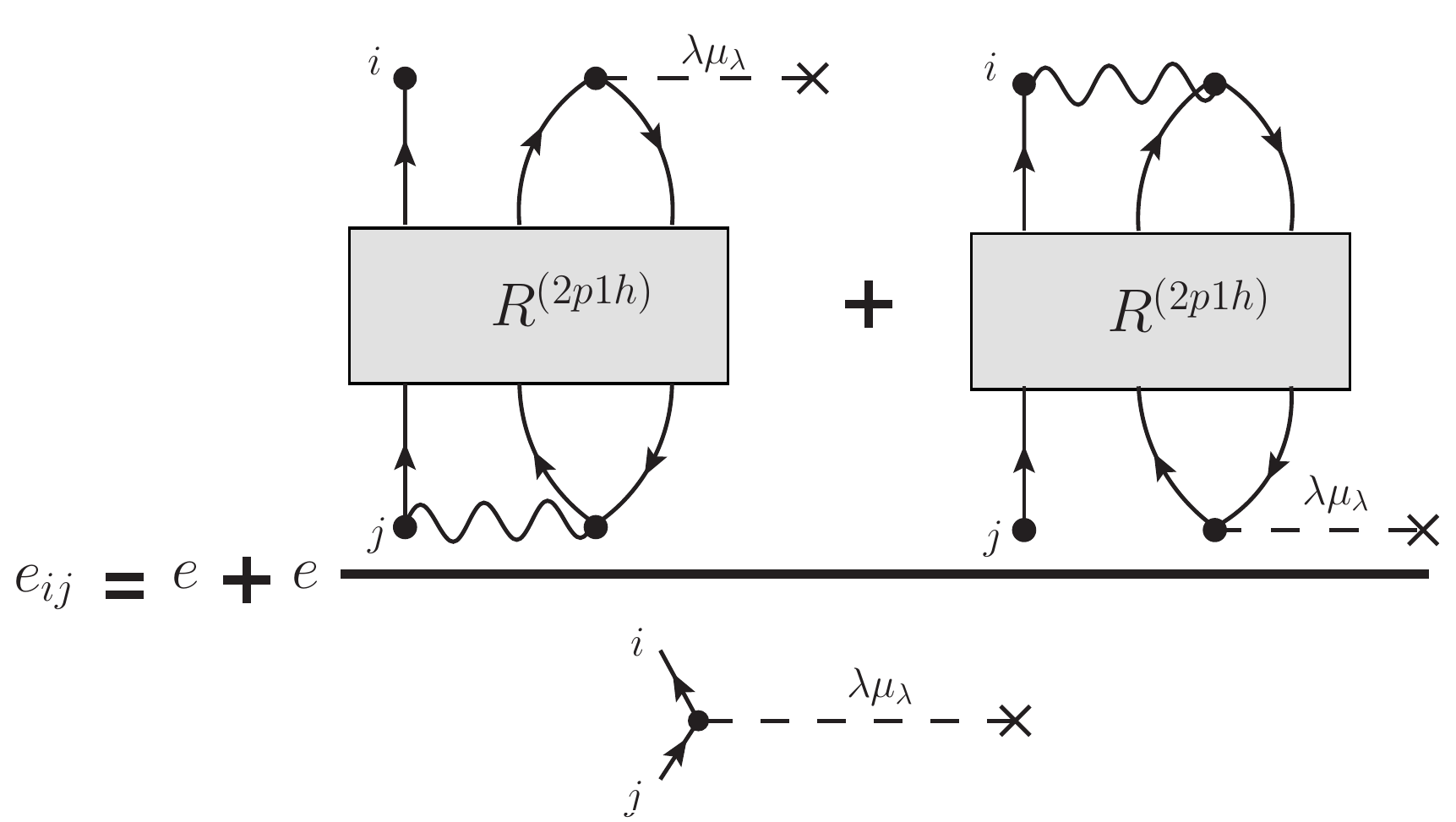}}
  \caption{ Schematic representation of the relation between the bare charge, denoted with $e$ (= 1 for protons, = 0 for neutrons), and the effective charge $e_{ij}$ required to renormalize the electromagnetic transition of multipolarity ($\lambda \mu_\lambda$). The ratio in the second term gives the correction to the bare charge due to the core-polarization effects.}
  \label{ratio_pict_V2}
\end{figure}

\begin{figure*}[t]
    \subfloat[]{\label{ADC3_a}\includegraphics[scale=0.5]{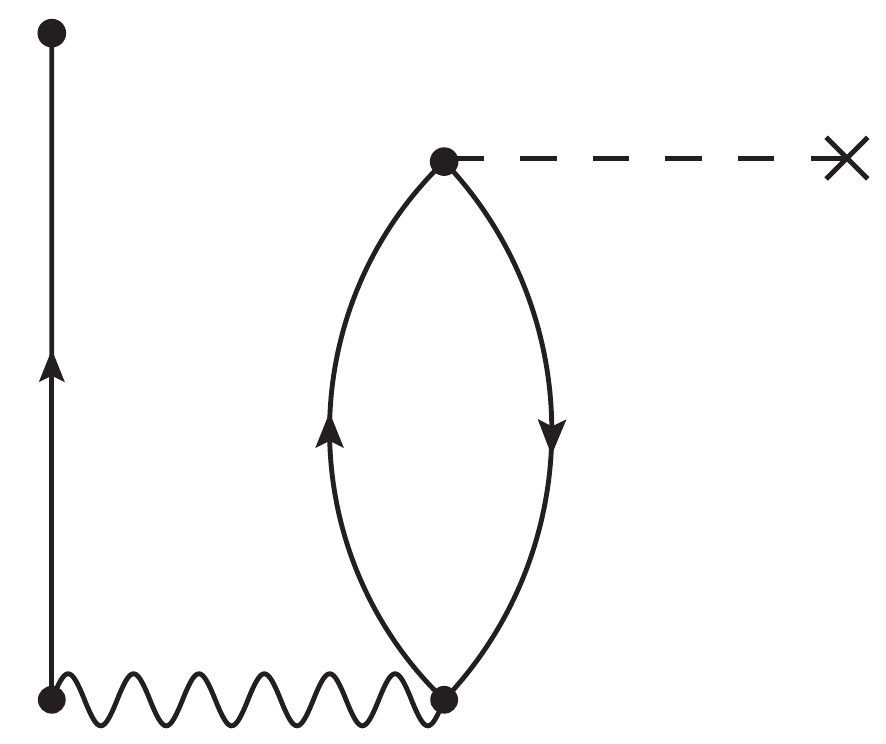}}
  \hspace{1.0cm}
    \subfloat[]{\label{ADC3_b}\includegraphics[scale=0.5]{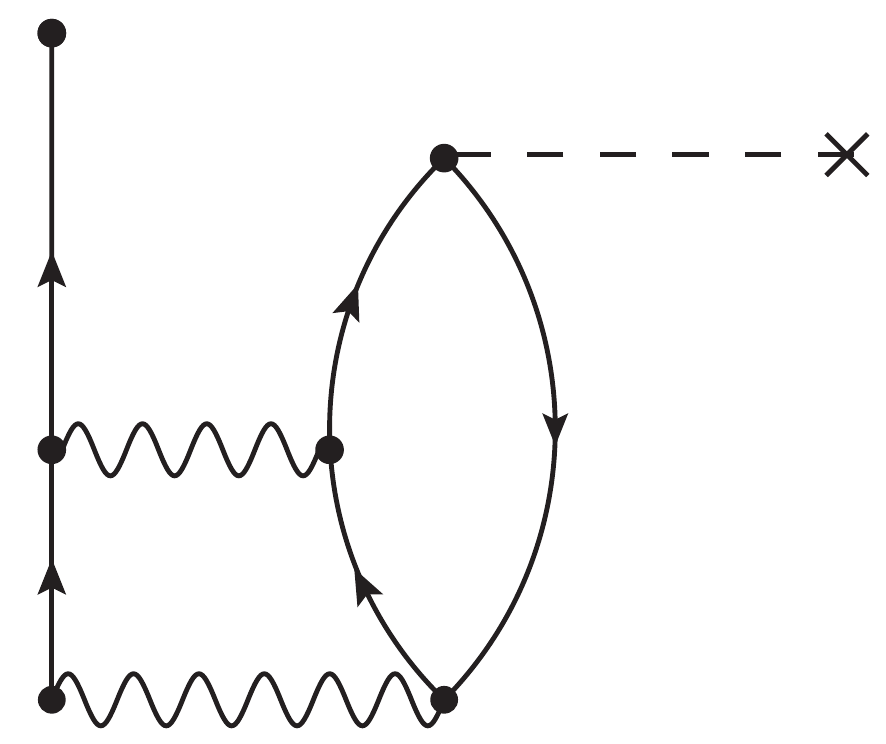}}
    \hspace{1.0cm}
    \subfloat[]{\label{ADC3_c}\includegraphics[scale=0.5]{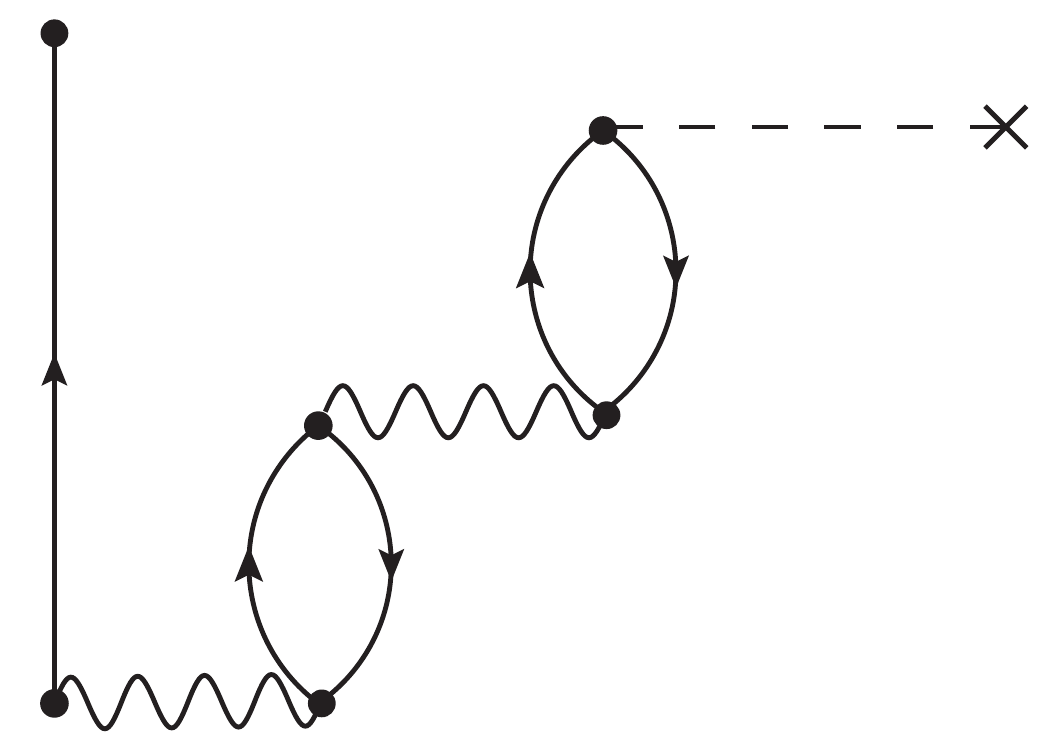}}
  \caption{Three different topologies of Feynman diagrams contributing to  $2p1h$ ISCs in Eq.~(\ref{eq:t_expansion}). The diagram~\ref{ADC3_a} is the only first-order contribution. Diagrams~\ref{ADC3_b} and~\ref{ADC3_c} are the second-order ones, containing fragments of ladder and a ring series respectively. }
  \label{ADC3_diagra}
\end{figure*} 

We note that there can be different ways to choose the energies and orbits~\eqref{eq:OpRS_MSP} that build the model space. A very appealing choice, from the physics point of view, is to select the dominant quasiparticle peaks that enter the exact propagator of Eq.~\eqref{Green_spectr}.  In this case, the orbits correspond to the very same single-particle-like excitations that have originally motivated the shell
 model~\cite{Mayer_PhysRev75_1949,Jensen_PhysRev75_1949}, 
 their wave functions would be transition amplitudes of Eq.~\eqref{tran_ampl} which describe one nucleon addition and removal processes, and the single-particle energies would correspond by construction to experimentally observed separation energies when only one valence nucleon is considered.  The major drawback of this route is that the transition amplitudes~\eqref{tran_ampl} are fragmented so that they are intrinsically normalised to have partial occupation, in contrast to the standard shell-model assumptions. This fragmentation can be generated in part from correlations inside the valence space and in part from outside, making it difficult to disentangle how much of this feature should be retained in the effective operator. Furthermore, whenever the experimental quasiparticle peaks are not well pronounced and rather fragmented, it becomes difficult to perform a one-to-one association with the single-particle space. Conversely, a set of mean-field orbits naturally defines an orthogonal single-particle basis and it poses no ambiguity on which states have to be identified as those belonging to the shell-model space. Among the many possible mean-field choices, the OpRS orbits~\eqref{eq:OpRS_MSP} are appealing since they are explicitly constructed to approximate as closely as possible both the exact quasiparticle energies and ground state properties that would be computed from the Dyson orbitals 
 of Eq.~\eqref{Green_spectr} ~\cite{Barbieri2009Ni56},~%
\footnote{The Hartree-Fock and the natural orbital bases are constructed to minimize the expectation value of the Hamiltonian and to maximize the overlap with the ground state wave function, under the constraint of a Slater determinant. However, they do miss important correlations. The OpRS basis can also be associated to a mean-field Hamiltonian but it aims at best approximating the correlated propagator~\eqref{Green_spectr}. The requirement of reproducing the lowest inverse moments of the spectral distribution ensures that the correlated nucleon density distribution and the total energy (through the Koltun sum rule) are reproduced
 closely~\cite{Barbieri2009Ni56}.}.

Once the valence space is set, we define the effective charge among two orbits $\ket{i}$ and $\ket{j}$ in terms of the ratio
\begin{equation}
e_{i j} = e \frac{\bra{\tilde i} \hat{\phi}^{(\lambda \mu_\lambda)}\ket{ \tilde j}} { \bra{i} \hat{\phi}^{(\lambda \mu_\lambda)}\ket{j} } \, ,
   \label{eq:def_ec}
\end{equation}
which is an orbital-dependent quantity, as apparent also from Fig.~\ref{ratio_pict_V2}. The denominator in Eq.~\eqref{eq:def_ec} is simply the bare matrix element of the electromagnetic operator between the valence space orbits:
\begin{align}
\bra{i} \hat{\phi}^{(\lambda \mu_\lambda)}\ket{j}  ={}&  \sum_{\alpha \beta} \; (z^i_\alpha)^\ast \, \phi^{(\lambda \mu_\lambda)}_{\alpha \beta}  \, z^j_\beta \;,
   \label{eq:rel_phi_VS}
\end{align}
where the matrix element of the external field $\phi^{(\lambda \mu_\lambda)}_{\alpha \beta} $ is calculated from the large model space for  the SCGF calculation (harmonic oscillator (HO) single-particle states  in this work).

The states $\ket{\tilde i}$ and $\ket{\tilde j}$ appearing in the numerator of Eq.~\eqref{eq:def_ec}  represent the correlated quasiparticle states that originate from the dynamical effects of coupling to phonons and other excitations that lie outside the valence space:
\begin{align}
\ket{ \tilde i} ={}&\ket{ i}  + \sum_{p , p' \notin S_V}  \sum_{\alpha} \Bigg[\frac{1}{\mathds{1} \epsilon^i - [\textbf{E}^{>}   +  \textbf{C}]}  \Bigg]_{\substack{\! \! p  p'}} \! \!  \textbf{M}_{p'  \alpha}  \, z^i_\alpha \,  \ket{p} \quad \quad ~
\nonumber \\
&+ \sum_{q , q' \notin S_V}  \sum_{\alpha}  \Bigg[ \frac{1}{ \mathds{1} \epsilon^i  - [\textbf{E}^{<}   +  \textbf{D}]}   \Bigg]_{\substack{\! \! q q'}} \! \! \!  \textbf{N}^{\dagger}_{q'  \alpha } \, z^i_\alpha  \,  \ket{q}  \, , 
   \label{eq:exp_i_tilde}
\end{align}
where the coupling and interactions matrices are the same as in Eq.~\eqref{irr_SE_Lehmann}  and the sum over the states $\ket{p}$ ($\ket{q}$) is limited only to those ISCs where at least one of the  $2p1h$ ($2h1p$) indexes lies outside $S_V$.

A few comments are in order in regard to Eq.~\eqref{eq:exp_i_tilde}: First, while this expression resembles the first order perturbation theory for the quasiparticle or quasihole states $\ket{\Psi^{A\pm1}_i}$%
\footnote{We recall that perturbation theory would give
\begin{align}
 {\ket {\tilde i}} ={}& {\ket i} + \sum_r \frac{\bra{r} \hat{H}_1 \ket{i} }{\epsilon^i - E^{>,<}_r}  {\ket r} \, ,   
\end{align}
where the configurations $r$ run only over $2p1h$~($2h1p$) states depending on the particle~(hole) character of ${\ket i}$ and ${\hat H}_1=-{\hat U}+{\hat V}+{\hat W}$ is the residual interaction}%
, it actually implies all-order resummations of the ISCs through the $\textbf{C}_{p p'}$ and $\textbf{D}_{q  q'}$ interaction matrices. The coupling terms $\textbf{M}_{p \alpha}$, $\textbf{N}_{\alpha q}$ also acquire contributions beyond first order as per the ADC(3) and FRPA formalisms.
Second,  in the spirit of the Dyson equation and the spectral content of the self-energy~\eqref{irr_SE_Lehmann},  contributions from both $2p1h$ and $2h1p$ are included.
 Third, the ISCs that belong to the valence space are suppressed from the  definition of $\ket{\tilde i}$. This is necessary since they will be diagonalised directly during the shell-model calculations and their contribution to the electromagnetic transition is already accounted for through the bare operator. To avoid double counting, they should not contribute to  the effective charges.
 Fourth, it should be recognized that denominators in Eq.~\eqref{eq:exp_i_tilde} could lead to unstable results whenever the choice of some single-particle energy $\epsilon^i$ is very close to a pole of the self-energy.  Since resummations of multi-particle-multi-hole configurations inside the valence space must not be included, the self-energy poles will generally be far from the valence space energies~$\epsilon^i$. Thus,  the chances for diverging denominators are greatly suppressed. However, we will see in Sect.~\ref{EffCH_calc} some cases when these poles cause instabilities.
Finally,  the expansion Eq.~\eqref{eq:exp_i_tilde} is in fact the same as the diagonalization of the Dyson equations. If we had chosen the $z^i_\alpha$  as the true quasiparticle amplitudes, then the $\ket{\tilde i}$ would be identified with the exact eigenvector that diagonalises the Dyson equation in matrix format
 (see ref.~\cite{Barb2017LNP}). 
 It is interesting to note that the coefficients in the above expansion are identified as components of a Dyson eigenvector which has finite norm. Thus, there cannot be divergences on the denominator above because of the overall normalization of $\ket{\tilde i}$.  This works adopts OpRS orbits for our valence space, which are only an approximation to the Dyson eigenvectors and energies, so instabilities cannot be ruled out exactly.

We then compute the matrix element of the external electromagnetic operator with respect to the quasiparticle state~(\ref{eq:exp_i_tilde}) up to linear terms in the $R^{(2p1h/2h1p)}(\omega)$ propagator:
\begin{widetext}
\begin{align}
\bra{\tilde i}& \hat{\phi}^{(\lambda \mu_\lambda)}\ket{\tilde j} = \bra{i} \hat{\phi}^{(\lambda \mu_\lambda)} \ket{j} \nonumber \\
   &+
     \sum_{\alpha \beta} \sum_{p \, p' \notin S_V}
     \left( z^i_\alpha  \right)^* \,               [\textbf{M}^{\bm\phi  (\lambda \mu_\lambda)}]^{\dagger}_{\alpha  p} 
           \Bigg[ \frac{1 } { \mathds{1} \epsilon^j -   [\textbf{E}^{>}   +  \textbf{C}]    } \Bigg]_{\substack{\! \! p  p'}} \! \!
          \textbf{M}_{ p'  \beta}   \, z^j_\beta
       +  \sum_{\alpha \beta} \sum_{q \, q' \notin S_V}           
           \left( z^i_\alpha  \right)^* \,          \textbf{N}^{\bm\phi  (\lambda \mu_\lambda)}_{\alpha \, q} 
              \Bigg[  \frac{1 }
              {  \mathds{1} \epsilon^j -     [\textbf{E}^{<}   +  \textbf{D}]   }  \Bigg]_{\substack{\! \! q q'}} \! \!
           \textbf{N}^\dagger_{q'  \beta}  \,    z^j_\beta
              \nonumber \\   
      &+ \sum_{\alpha \beta} \sum_{p \, p' \notin S_V}
      \left( z^i_\alpha \right)^{*} \, \textbf{M}^\dagger_{\alpha  p}
                 \Bigg[  \frac{1}
              { \mathds{1}  \epsilon^i - [\textbf{E}^{>}   +  \textbf{C}] }  \Bigg]_{\substack{\! \! p  p'}} \! \!
              \textbf{M}^{\bm\phi  (\lambda \mu_\lambda)}_{p' \, \beta}   \,    z^j_\beta
              +  \sum_{\alpha \beta} \sum_{q \, q' \notin S_V}  
                \left( z^i_\alpha \right)^{*} \,        \textbf{N}_{\alpha \, q}      \Bigg[    \frac{1 }
              {  \mathds{1} \epsilon^i -  [\textbf{E}^{<}   +  \textbf{D}] }    \Bigg]_{\substack{\! \! q q'}} \! \!    
               [\textbf{N}^{\bm\phi  (\lambda \mu_\lambda)}]^\dagger_{q'  \beta}   \,    z^j_\beta
                 \nonumber \\     
&=  \sum_{\alpha \beta} \; (z^i_\alpha)^\ast \, \Bigg\{
    \phi^{(\lambda \mu_\lambda)}_{\alpha \beta}  +   
  \widetilde\Sigma^{\text{L} (\lambda \mu)}_{\alpha\beta}(\omega =\epsilon^j  ) + \widetilde\Sigma^{\text{R} (\lambda \mu)}_{\alpha\beta}(\omega = \epsilon^i) 
  \Bigg\}  \, z^k_\beta \; .
\label{eq:t_expansion}
\end{align}
\end{widetext}
The last line of Eq.~\eqref{eq:t_expansion} shows that the corrections to the bare matrix element can be cast in a spectral representation analogous to the one of the dynamic self-energy where, however, the coupling of single-particle states to the $2p1h$ and $2h1p$ ISCs is now affected by the external field operator. We display analytical expressions for these coupling matrices ($\textbf{M}^{\bm\phi  (\lambda \mu_\lambda)}$ and $\textbf{N}^{\bm\phi  (\lambda \mu_\lambda)}$) in Eqs.~(\ref{eq:def_Mphi}-\ref{eq:def_Nphi}) below.

\begin{figure}[t]
    \subfloat[]{\label{RPA_pp}\includegraphics[scale=0.4]{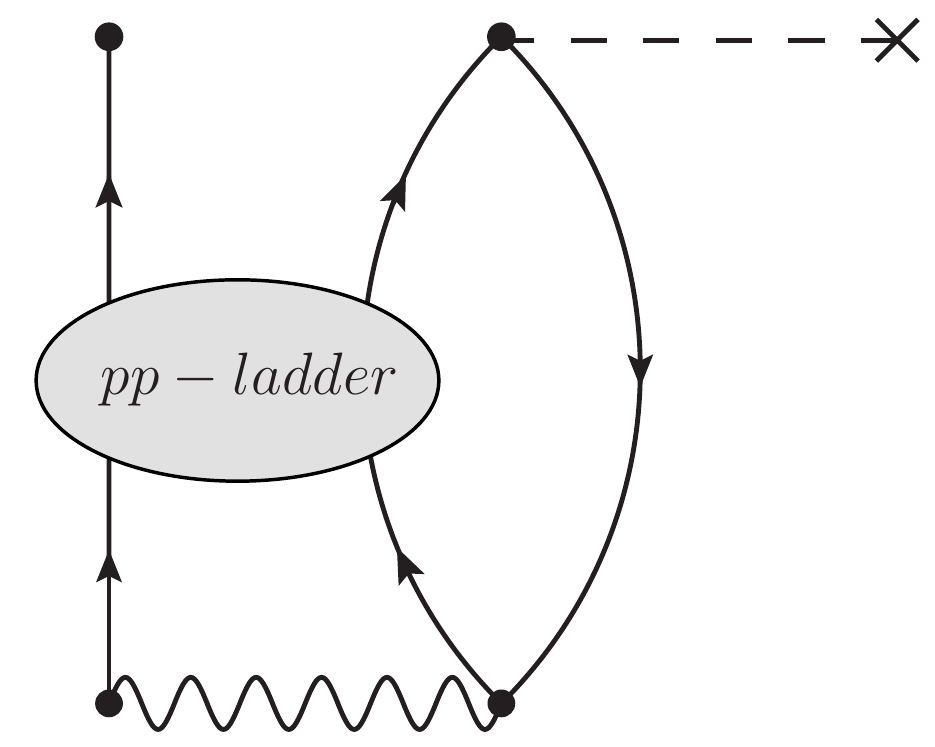}}
  \hspace{0.5cm}
    \subfloat[]{\label{RPA_ph}\includegraphics[scale=0.4]{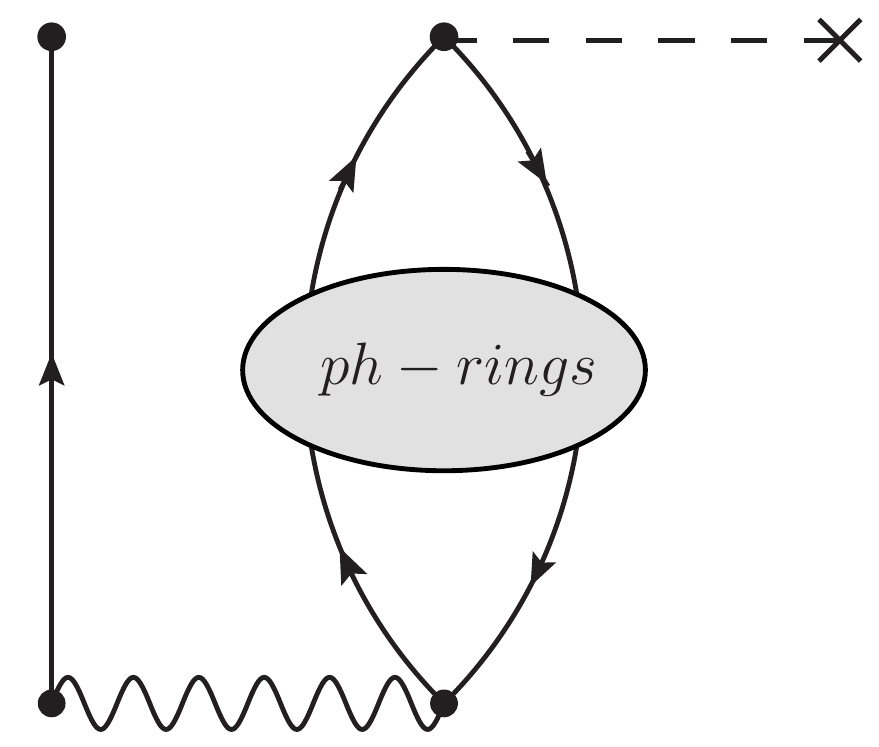}}
  \caption{
  Diagrammatic contributions to the effective charges that include at least one fully  resummed phonon of (a) ladder type or (b) ring type. The diagrams of Figs.~\ref{ADC3_b} and~\ref{ADC3_c} are contained in these contributions (\ref{RPA_pp} and \ref{RPA_ph}, respectively).}
  \label{RPA_diagra}
\end{figure}

\begin{figure}[t]
  \centering
    \subfloat{\includegraphics[scale=0.50]{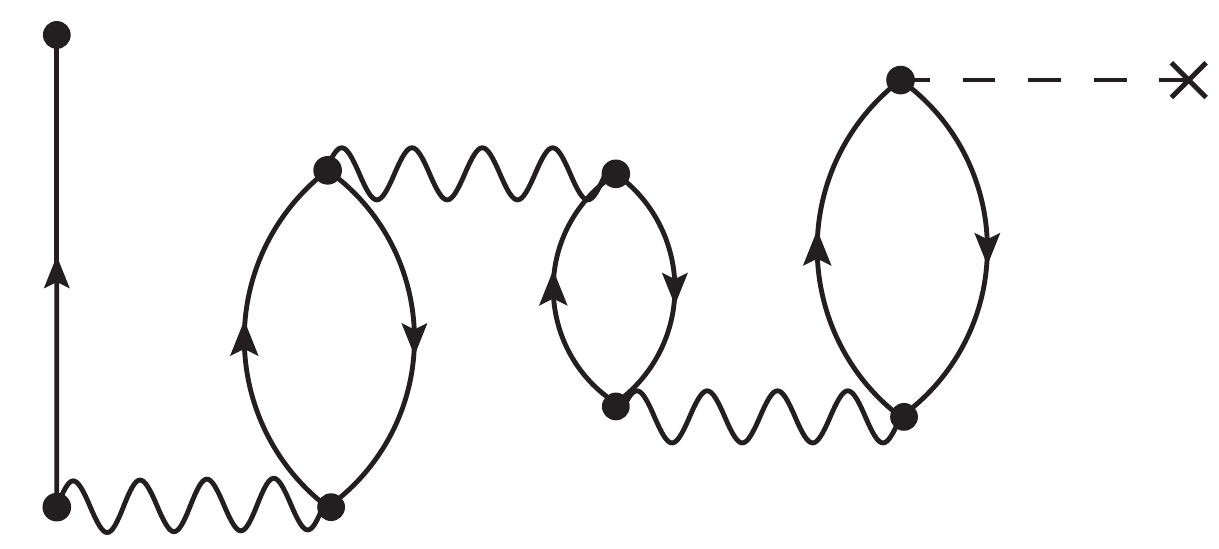}}
  \caption{Example of a diagram that is included in Fig.~\ref{RPA_ph}
  when using the FRPA expansion but that is neglected in the ADC(3) case. }
  \label{ADC3_c_RPA}
\end{figure}

Eqs.~\eqref{eq:def_ec} and~\eqref{eq:t_expansion} can also be represented by Feynman diagrams as shown in Fig.~\ref{ratio_pict_V2}. Since we calculate the  propagator $R^{(2p1h/2h1p)}(\omega)$ in the ADC(3) and FRPA approaches, it includes the nucleon-phonon coupling as discussed at the end of Sec.~\ref{form}. 
The lowest orders in the expansion of this propagator give rise to the diagrams of Fig.~\ref{ADC3_diagra} from where the $pp$/$hh$ (diagram \ref{ADC3_diagra}b) and the $ph$/$hp$ (\ref{ADC3_diagra}c) structures are evident. The corresponding contributions that include  the  resummation of one of these two types of phonon are  shown in Fig.~\ref{RPA_diagra}.  Diagram \ref{RPA_ph} is particularly relevant since it explicitly incarnates the phonon-mediated interaction of the external field with the valence particle. This is the phenomenological contribution originally 
discussed in~\cite{bohr1998} and
 it leads to the microscopic effective charges calculated in
 Refs.~\cite{Brown1974,Sagawa1979,SAGAWA198484}. 
 In the present work we employ the ADC(3) and FRPA frameworks to include ladders as in diagram \ref{RPA_pp} and to resum interference terms among ladders and rings to all orders (see Fig.~\ref{FRPA_exp}).
The main difference between the two approaches is that the ADC(3)/FTDA approach only allows for propagations of $pp$, $hh$ and $ph$ in one time direction, while the RPA also allows time inversion.  The latter generates additional time ordering such as the one shown in Fig.~\ref{ADC3_c_RPA}, which correspond to effectively accounting for $2p2h$ correlations in the ground state (on top of the OpRS ansatz)~\cite{RingSchuck}.

\begin{figure}[t]
  \centering
    \subfloat[]{\label{phi_left}\includegraphics[scale=0.55]{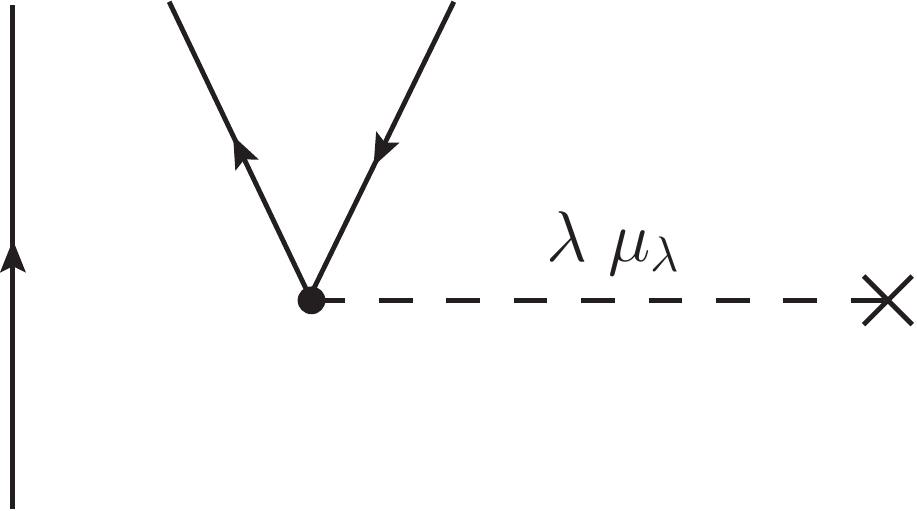}}
  \vspace{0.4cm}
   \subfloat[]{\label{phi_left_k}\includegraphics[scale=0.55]{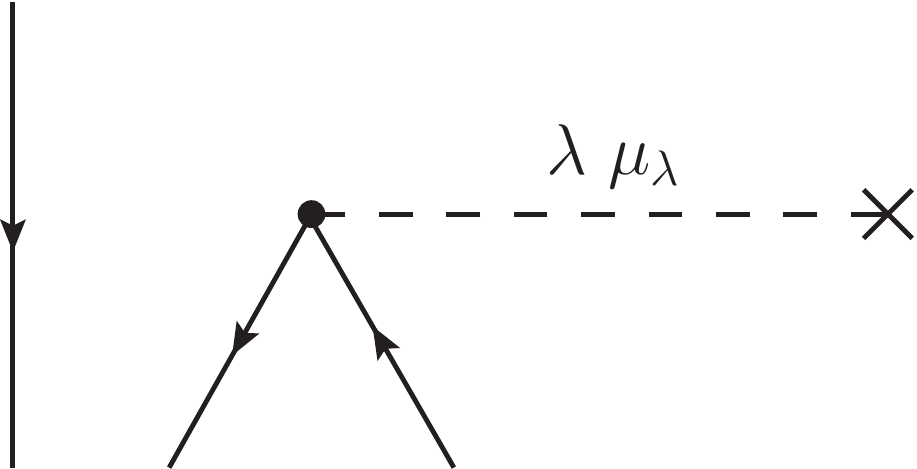}}
  \caption{Diagrammatic representation of the external field coupling vertices $\textbf{M}^{\bm\phi  (\lambda \mu_\lambda)}$ (\ref{phi_left}) and $\textbf{N}^{\bm\phi  (\lambda \mu_\lambda)}$ (\ref{phi_left_k})}
  \label{phi_A}
\end{figure}

The explicit expression for computing the $\textbf{C}_{p  p'}$, $\textbf{D}_{q  q'}$, $\textbf{M}_{p \alpha}$ and $\textbf{N}_{\alpha q}$ matrices, as well as details of how to solve the FRPA equation are reported in Refs.~\cite{Barb2017LNP,Soma2014,Barbieri2001frpa,Barbieri2007Ne,Degroote2011mols}.
In the above derivation we have also introduced two additional coupling matrices that link single-particle states  to the $2p1h$ and $2h1p$ ISCs by means of the external field. Their lowest order expressions are given by
\begin{equation}
  \textbf{M}^{\bm\phi  (\lambda \mu_\lambda)}_{p   \alpha} ~=~  \frac 1 {\sqrt{2}} \sum_{\gamma \,  \delta}\left( {\cal X}^{n_1}_{\alpha} {\cal X}^{n_2}_{\gamma} -{\cal X}^{n_2}_{\alpha} {\cal X}^{n_1}_{\gamma} \right) {\cal Y}^{k_3}_{\delta} ~  \phi_{\gamma \delta}^{(\lambda \mu_\lambda)} \, ,
   \label{eq:def_Mphi}
\end{equation}
with $p\equiv(n_1,n_2,k_3)$ for the forward-in-time part, and
 \begin{equation}
 \textbf{N}^{\bm\phi  (\lambda \mu_\lambda)}_{\alpha  q} ~=~ \frac 1 {\sqrt{2}} \sum_{\gamma \,  \delta} \phi_{\delta \gamma}^{(\lambda \mu_\lambda)} ~
  {\cal X}^{n_3}_{\delta}
 \left( {\cal Y}^{k_1}_{\alpha} {\cal Y}^{k_2}_{\gamma} - {\cal Y}^{k_2}_{\alpha} {\cal Y}^{k_1}_{\gamma} \right)
  \label{eq:def_Nphi}
\end{equation}
with $q\equiv(k_1,k_2,n_3)$ for the backward-in-time diagrams. These are displayed as fragments of Feynman/Goldstone diagrams in Figs.~\ref{phi_A}a and ~\ref{phi_A}b, respectively.  
The expressions are given in terms of the spectroscopic amplitudes~(\ref{tran_ampl}) of a dressed propagator as needed in the most general case of full self-consistency. In this work, however, we used the `\emph{sc0}' approximation 
of Ref.~\cite{Soma2014} in 
terms of the OpRS propagator, as it is the case for all state of the art applications of SCGF to finite nuclei.  
Note that Eqs.~(\ref{eq:def_Mphi}-\ref{eq:def_Nphi}) involve only the external field and are still inconsistent with the $\textbf{M}_{p \alpha}$ and $\textbf{N}_{\alpha q}$, for which ADC(3) and FRPA require corrections up to the second order in the residual interaction. Further improvements to this formalism will require deriving and implementing analogous corrections to the external field couplings.

%

\section{\label{EffCH_calc} Results for the Oxygen and Nickel chains}

Our purpose is to calculate shell-model effective charges by starting from realistic nuclear interactions, without relying on phenomenological comparisons with electromagnetic observables. We employ  the definition in Eq.~(\ref{eq:def_ec}), which is a  ratio between computed matrix elements of a given multipole field, with and without including correlations from outside the valence space.
 In this way we assess the impact of the core-polarization effects within the approximations discussed in Sec.~\ref{form2}.

We have considered medium-mass isotopes in the Oxygen ($^{14}\text{O}$, $^{16}$O, $^{22}$O and $^{24}$O) and Nickel ($^{48}$Ni, $^{56}$Ni, $^{68}$Ni and $^{78}$Ni) chains, corresponding to the nuclear shell-model valence spaces  $S_V=0p\, 1s\, 0d$ and $1p\,0f\,0g_{\frac{9}{2}}$, respectively. Effective charges for these two sets of isotopes (along with the chosen valence spaces) are taken as representative of their region in the nuclear chart, in analogy with the valence-space Hamiltonian operators in the 
nuclear shell-model~\cite{RevModPhys.77.427} and non-perturbative nuclear structure methods~\cite{Stroberg2017}.  

We choose the   chiral interaction 
NNLO$_{\text{sat}}$~\cite{Ekstrom2015} as 
microscopic nuclear  Hamiltonian with 2NF plus 3NF, where the three-body sector has been included in terms of an effective 
2N interaction~\cite{Cipol13}. 
The HO basis in the calculations is truncated at $N_{\text{max}}$~=~13 (14 major shells) for the Oxygen isotopes and at $N_{\text{max}}$~=~11 for Nickel (due to computational limits); in all cases, an HO frequency of \hbox{$\hbar\Omega$~=~20 MeV} has been used.
We focus on the quadrupole isoscalar effective charges and we use the corresponding electromagnetic  E2 operator, 
 \begin{equation}
\hat{\phi}^{(2 \mu)} = \sum_i r_i^2 \, Y_{2 \mu}(\hat{r}_i),
  \label{eq:quad_oper}
\end{equation}
in Eqs.~(\ref{eq:def_Mphi}-\ref{eq:def_Nphi}).

As explained  in Sec.~\ref{form2}, our calculations are based on the OpRS propagator that is obtained by solving the Dyson equation with the ADC(3) self-energy; likewise, the correction terms $\widetilde\Sigma^{\text{L} (\lambda \mu)}_{\alpha\beta}(\omega)$ and $\widetilde\Sigma^{\text{R} (\lambda \mu)}_{\alpha\beta}(\omega)$ in Eq.~\eqref{eq:t_expansion} also contain the ADC(3) or FRPA expansion of  $R^{(2p1h/2h1p)}(\omega)$,  meaning that the diagrams topologies in Fig.~\ref{ADC3_diagra}  are used as the \enquote{seeds} for  all-orders resummations of phonons as those entering the diagrams of Fig.~\ref{RPA_diagra}. 
We have found that the difference in the charge renormalization computed in the FTDA and in the FRPA is within the 20\% in the majority of cases. 
For a few instances, we observe an erratic behaviour of the computed effective charges  that is related to the divergences of the poles in Eq.~\eqref{eq:exp_i_tilde}, as discussed in Sec.~\ref{form2}.  We will investigate some specific cases further below, in which the energy difference in these denominators can be of the order of few KeV and it triggers unphysical  values of the  effective charges in Eq.~(\ref{eq:t_expansion}).   In spite of these few anomalies,  we find that trends in the computed charges are still visible.  
This issue arises because we exploit the true poles of the self-energy in Eqs.~(\ref{eq:exp_i_tilde}) but we approximate the quasiparticle energies with the poles  $\epsilon^i$ of the $g_{\alpha\beta}^{\rm{OpRS}}$ propagator,  when the exact Dyson eigenstates would instead be required for consistency.
To identify these pathological cases,  we have investigated the behaviour of the energy denominators  in Eq.~(\ref{eq:t_expansion}) every time our calculations produced values of the effective charge that appeared to be unphysical.

\subsection{\label{Oxygen}  Oxygen isotopes in the $0p\, 1s\, 0d$ valence space} 

We first discuss  E2 effective charges for the $^{14}$O, $^{16}$O, $^{22}$O and $^{24}$O isotopes.
The single-particle energies and orbits forming the  valence space are those of the $g_{\alpha\beta}^{\rm{OpRS}}$ propagator~(\ref{g_OpRS}) and  we consider the $0p_{\frac{1}{2}}$, $0p_{\frac{3}{2}}$, $0d_{\frac{3}{2}}$, $0d_{\frac{5}{2}}$ and $1s_{\frac{1}{2}}$ states for both neutrons and protons.  The $e_{\nu}$ and $e_{\pi}$ calculated in FRPA  for the four isotopes are collected in Tables~\ref{tab:Oxy_Neutron} and~\ref{tab:Oxy_Proton}, respectively. The tables also display the FTDA values in parentheses whenever the deviations from the bare charges differ more than 10\% form the corresponding FRPA results.

\begin{table}[b]
\caption{\label{tab:Oxy_Neutron} Isoscalar E2 $e_{\nu}$ of $^{14}$O,  $^{16}$O, $^{22}$O and $^{24}$O for \hbox{$S_V=0p\, 1s\, 0d$}. The virtual phonons coupling with the external field are calculated in FRPA. The values obtained in the FTDA are also  shown in parentheses, whenever they differ from the FRPA ones by more than 10\%. The results are calculated in an HO model space with $N_{\text{max}}$=~13  and $\hbar\Omega$ = 20 MeV. }
\begin{ruledtabular}
\begin{tabular}{c||c|c|c|c}
&  $^{14}$O & $^{16}$O  & $^{22}$O &  $^{24}$O    \\
\hline
$ s_{\frac{1}{2}}$ $ d_{\frac{3}{2}}$  & 0.27 &  0.20 & 0.12 & 0.14  (0.12) \\
$ s_{\frac{1}{2}}$ $ d_{\frac{5}{2}}$  &   0.42      &  0.30      &   0.23      &       0.26     \\
$ p_{\frac{1}{2}}$ $ p_{\frac{3}{2}}$  &   0.45 (0.41)     &     0.48     &    0.56 (0.41)    &     0.45       \\
$ p_{\frac{3}{2}}$  &   0.43  (0.48)    &     0.30 (0.36)     &     0.48 (0.95)   &       0.38 (0.32)     \\
 $ d_{\frac{3}{2}}$  &    0.28     &    0.20      &    0.16     &      0.17      \\
$ d_{\frac{3}{2}}$ $ d_{\frac{5}{2}}$  &    0.47     &    0.37      &    0.26     &    0.26  (0.23)      \\
 $ d_{\frac{5}{2}}$  &    0.45     &    0.304    &   0.34      & 0.32           \\
\end{tabular}
\end{ruledtabular}
\end{table}

\begin{figure}[t]
  \centering
    \subfloat{\includegraphics[scale=0.33]{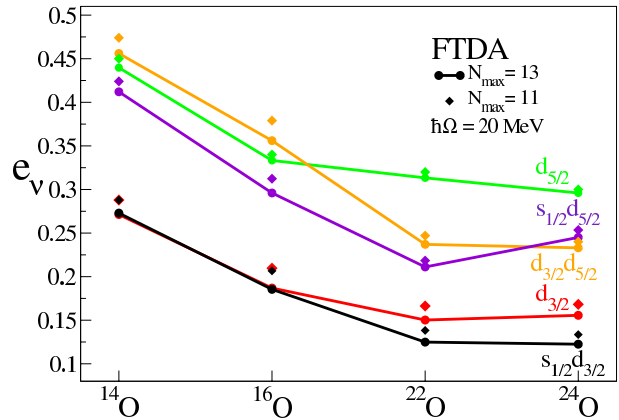}}
  \caption{Oxygen isotopes $e_\nu$  computed in FTDA. Values with circles (diamonds) are calculated in an $N_{\text{max}}$ = 13 (11) HO model space. Straight lines connecting the values at $N_{\text{max}}$=~13  are meant to guide the eye.}
  \label{TDA_conv_13_11}
\end{figure}

\begin{table}[b]
\caption{\label{tab:Oxy_Proton} Same as in Table~\ref{tab:Oxy_Neutron} but for proton effective charges. Both FRPA and FTDA numbers are shown if the corrections from the bare charge ($e_\pi$=1) differ from each other by more than 10\%.}
\begin{ruledtabular}
\begin{tabular}{c||c|c|c|c|c|c|c|c}
 &  $^{14}$O & $^{16}$O  & $^{22}$O &  $^{24}$O    \\
\hline
$ s_{\frac{1}{2}}$ $ d_{\frac{3}{2}}$  & 0.87 (0.69) & 1.067 (1.058) & 1.04 & 1.04 (1.03)  \\
$ s_{\frac{1}{2}}$ $ d_{\frac{5}{2}}$  &    1.18    &   1.14      & 1.16       &  1.16        \\
$ p_{\frac{1}{2}}$ $ p_{\frac{3}{2}}$  &    1.18     &   1.15     &    1.23     &     1.18      \\
$ p_{\frac{3}{2}}$  &    1.02     &   1.001 (1.014)     &   1.09 (1.07)    &    1.06 (1.05)     \\
 $ d_{\frac{3}{2}}$  &   1.00 (0.46)     &    1.018 (1.033)     &   1.03 (1.04)     &    1.01  (1.03)     \\
$ d_{\frac{3}{2}}$ $ d_{\frac{5}{2}}$  &   1.04 (0.79)     &    1.139 (1.158)     &    1.23    &      1.18      \\
 $ d_{\frac{5}{2}}$  &   1.14   &    1.08       &   1.13  (1.11) &    1.09      \\
\end{tabular}
\end{ruledtabular}
\end{table}
For the neutron effective charges, we see a significant dependence on the quasiparticle orbitals considered, especially for the $^{22}$O, spanning a wide range of  values, from $e_{\nu \!  s_{\! \frac{1}{2}} \, \! \nu \! d_{\! \frac{3}{2}}}$=~0.12  to $e_{\nu \! p_{\!\frac{1}{2}} \, \! \nu \! p_{\! \frac{3}{2}}}$=~0.56. The $e_{\nu}$ values involving the $d_{\frac{3}{2}}$ orbital are rather low ($<$ 0.3 in most cases) also for the isotopes in the valley of stability, $^{14}$O and $^{16}$O. 
 In order to check the convergence our calculations, we compare the $e_\nu$ values at  $N_{\text{max}}$ = 13 with the ones computed in a smaller $N_{\text{max}}$ = 11 space. This is done at the FTDA level and it is shown in Fig.~\ref{TDA_conv_13_11}. Overall, we find good convergence, indicating that  observed trends in the computed charges are not largely affected by the truncation of the large HO model space.  The $d_{\frac{3}{2}}$ effective charge shows the largest variations and this is also accompanied by a  slow convergence of the OpRS single-particle energy for this orbital.

\begin{figure}[t]
  \centering
    \subfloat{\includegraphics[scale=0.33]{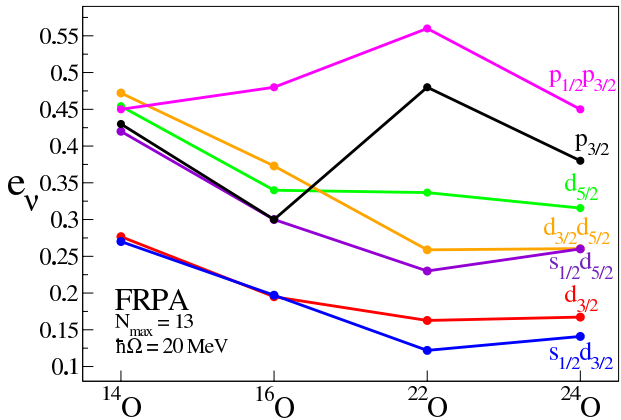}}
  \caption{Oxygen isotopes  $e_\nu$ computed in FRPA  (see Table~\ref{tab:Oxy_Neutron}). 
  Straight lines are a guide to the eye.}
  \label{e_n_plot}
\end{figure}
The four isotopes considered here  go from the proton-rich $^{14}$O 
to the neutron drip line. Neutron effective charges  show a distinctive isotopic trend, decreasing with the number of neutrons, as displayed also in Fig.~\ref{e_n_plot}. Most values of $e_\nu$ in  $^{24}$O are about half as compared to the ones in $^{14}$O. The only exception to this trend is given by the $p$-shell orbits  in Fig.~\ref{e_n_plot}, where the $e_\nu$ appear to be quite insensitive to the isotopic changes of many-body correlations. This is expected since these are strongly bound orbits in the single-particle spectrum.
The exception to this flat trend  is given by the $p$-shell $e_{\nu}$ of $^{22}$O that follows an irregular pattern. The reason for this is related to the
raise of degeneracies in the denominators of Eq.~\eqref{eq:t_expansion}  and it is discussed below. 
 A closer comparison between the $e_\nu$ values for $^{22}$O and $^{24}$O reveals a flat trend near the dripline, with the $^{24}$O  charges slightly bigger than the $^{22}$O ones in a few cases. By comparing the values of point-neutron radii $\langle r_{\nu}^{2} \rangle^{1/2}$ of the four Oxygen isotopes in Table~\ref{Tab:Radii}, we see that this may be due to the saturation of the polarization effect, as the two extra nucleons in the $^{24}$O do not change drastically the neutron distribution in the isotope. 
Note that we do not find any sizeable deterioration of the convergence with respect to the HO space for $^{22}$O and $^{24}$O  in Fig.~\ref{TDA_conv_13_11}, compared to the other Oxygen isotopes, that could be responsible for this inversion.
\begin{table}[b]
\caption{\label{Tab:Radii} Theoretical point-neutron and point-proton intrinsic radii (in fm) of $^{14}$O,  $^{16}$O, $^{22}$O and $^{24}$O, computed in at $N_{\text{max}}$=~13  and $\hbar\Omega$=~20 MeV. }
\begin{ruledtabular}
\begin{tabular}{c||c|c|c|c}
&  $^{14}$O & $^{16}$O  & $^{22}$O &  $^{24}$O    \\
\hline
 $\langle r_{\nu}^{2} \rangle^{1/2}$ & 2.37 & 2.62  & 2.93 & 3.11 \\
$\langle r_{\pi}^{2} \rangle^{1/2}$  & 2.57 & 2.64  & 2.67 & 2.71  \\
\end{tabular}
\end{ruledtabular}
\end{table}

\begin{figure}[t]
  \centering
    \subfloat{\includegraphics[scale=0.33]{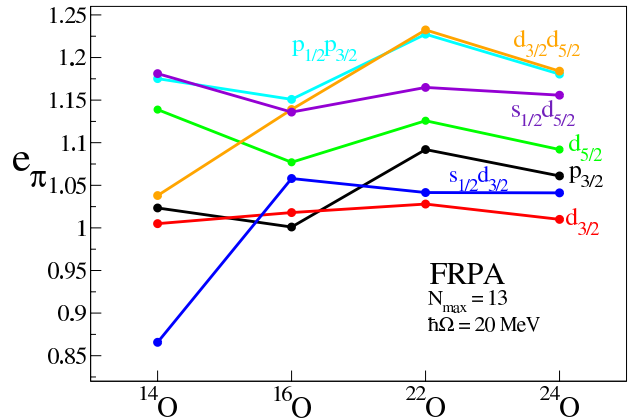}}
  \caption{ Oxygen isotopes $e_\pi$ computed in FRPA (see Table~\ref{tab:Oxy_Proton}).  Straight lines are a guide to  the eye.}
  \label{e_p_plot}
\end{figure}
 Only FTDA charges differing by more than 10\% compared to the FRPA ones are shown in Tables~\ref{tab:Oxy_Neutron}-~\ref{tab:Oxy_Proton}, while in the other cases the inclusion of the FRPA phonons does not change significantly the renormalization of the charge. For $e_\nu$, the $ p_{\frac{3}{2}}$ orbital is particularly sensitive to the approximation used to compute the intermediate phonon excitations, especially in $^{22}$O where the FTDA overestimates the polarization effect on the charge. 
 In both the FTDA and FRPA cases we see an erratic trend of the charges associated to the $p_{\frac{3}{2}}$ neutron orbital, which is apparent from the $p_{\frac{3}{2}}$  and $p_{\frac{1}{2}} p_{\frac{3}{2}}$ curves in Fig.~\ref{e_n_plot}. Table~\ref{tab:Neutron_p3_2} displays the quasiparticle energies $\epsilon^i$ for the neutron $p_{\frac{3}{2}}$ orbit and the  pole of the self-energy, $\varepsilon^\Sigma$, that is closest to it, for each of the isotopes under consideration.  The poles of $\widetilde{\Sigma}(\omega)$ are found by direct diagonalization of the matrices $[\textbf{E}^{>}   +  \textbf{C}]$ and $[\textbf{E}^{<}   +  \textbf{D}]$ in Eq.~\eqref{irr_SE_Lehmann}. For the  $^{22}$O, the unnaturally large $e_{\nu p_{\frac{3}{2}}}$ in FTDA corresponds to the presence of a vanishing denominator (down to a few keV) in Eq.~\eqref{eq:t_expansion}.
\begin{table}[b]
\caption{\label{tab:Neutron_p3_2} Quasiparticle energies  from the OpRS propagator  (second column), poles of the self-energy computed in FTDA (third column) and FRPA (fifth column) and  effective charges in FTDA (fourth column) and FRPA (sixth column) of the $p_{\frac{3}{2}}$ neutron  orbital for the Oxygen isotopes. Energies are given in MeV.}
\begin{ruledtabular}
\begin{tabular}{c||c|c|c|c|c}
&                        &  \multicolumn{2}{c|}{FTDA}  &   \multicolumn{2}{c}{FRPA}  \\
& $\epsilon^{\nu p_{\frac{3}{2}}}$  & $\varepsilon^\Sigma$  & $e_{\nu p_{\frac{3}{2}}}$  &  $\varepsilon^\Sigma$  &  $e_{\nu p_{\frac{3}{2}}}$  \\
\hline
$^{14}$O   &  -23.72     &  -34.92  &  0.48   &    -31.10   &   0.43  \\
 $^{16}$O  &  -24.44     &  -32.83    &   0.36    &   -30.32     &  0.30   \\
$^{22}$O  &  -17.8156    & -17.8103   &   0.95    &  -17.54      &  0.48   \\
 $^{24}$O &   -21.46    &   -21.57  &   0.32   &   -21.35     &   0.38  \\    
\end{tabular}
\end{ruledtabular}
\end{table}

In general our results are in line with the typical values required to match experimental electric quadrupole properties for $0p\, 1s\, 0d$ nuclei:  a quench of the neutron effective charges standard values has been found for neutron-rich isotopes of Boron, Carbon, Nitrogen, Oxygen and 
Neon~\cite{Hamamoto1996,Sagawa2004,Yuan2012}. Since 
the pioneeristic study by Bohr and Mottelson~\cite{bohr1998}, this 
effect is explained as the result of the decoupling of the valence neutrons from the protons in the core, and indeed we see a decreasing trend in the charge values for increasing number of neutrons in the Oxygen chain.

As compared to the neutron effective charges, the orbital dependence is less pronounced in the $e_\pi$ shown in Table~\ref{tab:Oxy_Proton}, where almost all the values are within \hbox{the 1.0-1.2 range}. These $e_\pi$ values are all consistently smaller than the standard phenomenological values of $\sim$1.5. The renormalization effect in the protons is weaker in magnitude than in the neutrons. Therefore, we find relative differences  bigger than 10\% between FTDA and FRPA for several cases of $e_\pi$.
In general, proton charges do not have an isotopic dependence, as shown by the relatively flat trends in Fig.~\ref{e_p_plot}. This is reflected by the slowly varying behaviour of the radial proton distribution, as apparent from the $\langle r_{\pi}^{2} \rangle^{1/2}$ values in Table~\ref{Tab:Radii}.

For the transition involving the  $\pi d_{\frac{3}{2}}$ orbitals of the neutron-deficient nucleus $^{14}$O, we obtain a 
negative correction to the proton charge, as one can see from the values $<$ 1. Also in this case, this particular value is invalidated by a small energy denominator in Eq.~(\ref{eq:t_expansion}) that generates an erratic behaviour. The details of the single-particle energies for  the $\pi d_{\frac{3}{2}}$ orbital  and of the self-energy poles are compared to the resulting effective charge in Table~\ref{tab:proton_d3_2} for various HO model spaces and frequencies. Again, unphysical values  of $e_\pi$ are computed when the energy denominators become smaller than $\approx$50~KeV, as for the FTDA calculation at  $N_{\text{max}}$=11 with $\hbar \omega$=18 MeV.
\begin{table}[b]
\caption{\label{tab:proton_d3_2} Quasiparticle energies from the OpRS propagator (second column), poles of the self-energy (third column) and effective charges  (fourth column) of the $d_{\frac{3}{2}}$ proton  orbital for a $N_{\text{max}}$=13 HO space with $\hbar \omega$=20 MeV, and for $N_{\text{max}}$=11 with $\hbar \omega$=18, 20 and 22 MeV in $^{14}$O. Energies are given in MeV and calculation are in FTDA unless explicitly stated.}
\begin{ruledtabular}
\begin{tabular}{c||c|c|c}
$N_{\text{max}}$, $\hbar \omega$ & $\varepsilon_{\pi d_{\frac{3}{2}}}$  & $\varepsilon^\Sigma$ & $e_{\pi d_{\frac{3}{2}}}$   \\
\hline
13, 20 MeV (FRPA)  &  7.608    & 7.127  &  1.00  \\
 13, 20 MeV   &  7.608    & 7.554  &   0.46  \\
 11, 18 MeV    &   7.7259   & 7.743  &  3.54  \\
  11, 20 MeV   &  8.1301    & 8.491  &  1.24  \\  
 11, 22 MeV   &    8.5427  &  9.298  &  1.18  \\    
\end{tabular}
\end{ruledtabular}
\end{table}

\subsection{\label{Nickel} Nickel isotopes in the $1p\,0f\,0g_{\frac{9}{2}}$ valence space}

The E2 neutron and proton effective charges for $^{48}$Ni, $^{56}$Ni, $^{68}$Ni and $^{78}$Ni isotopes presented in this section were computed for $S_V=1p\,0f\,0g_{\frac{9}{2}}$.

In the shell-model literature on $pf$ shell nuclei, E2 effective charges of $e_\nu \simeq $ 0.5 and $e_\pi \simeq $ 1.5 are referred to as the \lq\lq standard\rq\rq  
values~\cite{POVES1981235}.  
The assumption is that the effect of polarization ($\delta e \simeq$ 0.5) is the same for both neutrons and protons. However, the measurement of the isoscalar and isovector polarization effects on mirror nuclei $^{51}$Fe and $^{51}$Mn in 
Ref.~\cite{duRietz2004}, suggested 
different values $e_\nu \simeq $ 0.80 and $e_\pi \simeq $ 1.15. Similarly, in other 
studies~\cite{Brown1974,Hoischen2011,Allmond2014,Loelius2016}, the 
above  standard charges  have been found to miss the agreement  with the experimental values of static moments and electromagnetic transitions for several $pf$-shell isotopes. Moreover, 
Ma \textit{et al.}~\cite{Ma2009} have applied the microscopic particle-vibration model with a Skyrme parametrization of the nuclear interaction, and they found the  standard values for $e_\nu$ but a quenched value of $e_\pi \simeq $ 1.3 for the $^{44, 46, 48}$Ti isotopes.

The  values of E2 $e_\nu$ and $e_\pi$ that we obtain for the four Nickel isotopes are collected in Tables~\ref{tab:Nickel_Neutron_RPA} and~\ref{tab:Nickel_Proton_RPA}, respectively. In general, the orbital dependence within each isotope is significant for $e_\nu$. For instance, the $^{48}$Ni nucleus computed within the FRPA spans a range of  values from $e_{\nu \! p_{\!\frac{3}{2}}}$=~0.46 to $e_{\nu \! f_{\!\frac{5}{2}} \, \!\nu f_{\!\frac{7}{2}}}$=~0.82. The orbital dependence is less pronounced for the $e_\pi$, in line with what we have seen for the Oxygen nuclei. 

\begin{table}[b]
\caption{\label{tab:Nickel_Neutron_RPA} Isoscalar E2 $e_{\nu}$ charge for $^{48}$Ni,  $^{56}$Ni, $^{68}$Ni and $^{78}$Ni assuming  $S_V=1p\,0f\,0g_{\frac{9}{2}}$.  The calculations are performed in FRPA, with  \hbox{$N_{\text{max}}$ = 11} and $\hbar\Omega$ = 20 MeV. The FTDA results that differ from the FRPA ones by 10\% or more are also displayed in parentheses.}
\begin{ruledtabular}
\begin{tabular}{c||c|c|c|c}
 &  $^{48}$Ni & $^{56}$Ni & $^{68}$Ni & $^{78}$Ni   \\
\hline
$ f_{\frac{5}{2}}$   & 0.61    & 0.54    &  0.40  &  0.47 \\
$ f_{\frac{5}{2}}$ $ f_{\frac{7}{2}}$     & 0.82    &   0.57 &  0.59  &   0.62 (0.39) \\
$ f_{\frac{5}{2}}$ $ p_{\frac{1}{2}}$       &  0.53   & 0.48 (0.44)  & 0.34   &  0.45 (0.39)   \\
$ f_{\frac{5}{2}}$ $ p_{\frac{3}{2}}$       & 0.57 (0.52)   & 0.49  (0.45)  &  0.37  &   0.53 (0.45)  \\
$ f_{\frac{7}{2}}$      &  0.57 & 0.48 (0.44)    &  0.36  & 0.27 (0.39)  \\
$ f_{\frac{7}{2}}$   $ p_{\frac{3}{2}}$    &  0.66  & 0.53 (0.48)   &   0.46  & 0.78 (-1.54)   \\
$ p_{\frac{1}{2}}$ $ p_{\frac{3}{2}}$  &  0.48 &  0.40  & 0.29   &  0.29  \\
$ p_{\frac{3}{2}}$       &  0.46   &  0.39   &   0.29  &   0.31  \\
$ g_{\frac{9}{2}}$       &  0.57 (0.52)   &  0.47  &  0.36   &  0.38    \\
\end{tabular}
\end{ruledtabular}
\end{table}

\begin{table}[b]
\caption{\label{tab:Nickel_Proton_RPA}  Same as in Table~\ref{tab:Nickel_Neutron_RPA} but for proton effective charges. Both FRPA and FTDA numbers are shown if the corrections from the bare charge ($e_\pi$=1) differ from each other by more than 10\%.}
\begin{ruledtabular}
\begin{tabular}{c||c|c|c|c}
 &  $^{48}$Ni & $^{56}$Ni & $^{68}$Ni & $^{78}$Ni   \\
\hline
$ f_{\frac{5}{2}}$   &  1.15   &  1.14  &  1.06 (1.07)  &  1.14 \\
$ f_{\frac{5}{2}}$ $ f_{\frac{7}{2}}$     &  1.14   & 1.15 (1.17)    &  1.09  &  1.04 (1.08)   \\
$ f_{\frac{5}{2}}$ $ p_{\frac{1}{2}}$       &  1.08 (1.07)  &  1.08 (1.07)  &  1.03 (1.02)  &  1.23 (1.14)   \\
$ f_{\frac{5}{2}}$ $ p_{\frac{3}{2}}$       & 1.09   &  1.09  &  1.04  &   1.28 (1.18)  \\
$ f_{\frac{7}{2}}$      &  1.15 &  1.12  &  1.05 (1.07)  &  1.14  \\
$ f_{\frac{7}{2}}$   $ p_{\frac{3}{2}}$    & 1.19   &  1.20  &  1.17  &  1.23  \\
$ p_{\frac{1}{2}}$ $ p_{\frac{3}{2}}$  & 1.11  & 1.09   &  1.05 (1.07)  &  1.14 (1.11)  \\
$ p_{\frac{3}{2}}$       &   1.10  & 1.07 (1.08)   &  1.04 (1.06)  &  1.13  \\
 $ g_{\frac{9}{2}}$           &  1.19   & 1.15   &  1.11  &   1.20   \\
\end{tabular}
\end{ruledtabular}
\end{table}

\begin{figure}[t]
  \centering
    \subfloat{\includegraphics[scale=0.33]{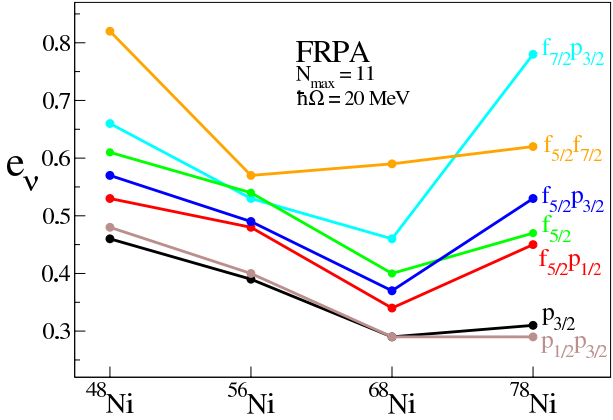}}
  \caption{ Values of $e_\nu$ for selected orbitals of  $^{48}$Ni, $^{56}$Ni, $^{68}$Ni $^{78}$Ni  and displayed in Table~\ref{tab:Nickel_Neutron_RPA}. The effective charges for $g_{\frac{9}{2}}$ and $f_{\frac{7}{2}}$  almost overlap with the ones for the $f_{\frac{5}{2}}p_{\frac{3}{2}}$ transition and they are not shown for clarity. Straight lines are a guide to the eye.}
  \label{e_n_plot_NICKEL}
\end{figure}

As shown in Table~\ref{tab:Nickel_Neutron_RPA}, the two doubly-magic nuclei $^{48}$Ni and $^{56}$Ni have $e_\nu$ around the standard values \hbox{$e_{\nu} \simeq $ 0.4-0.6}. The transition within the $f_{\frac{5}{2}}$ and $f_{\frac{7}{2}}$  orbitals seems to be more affected by polarization effects, as indicated by the larger charge of  $e_\nu = $ 0.82 in $^{48}$Ni. When we move to $^{68}$Ni, we observe a similar trend as in the neutron-rich Oxygen isotopes, with a quench of the polarization effect that brings the neutron effective charges around the values of $e_\nu \simeq $ 0.3-0.4. This trend is better visualized in Fig.~\ref{e_n_plot_NICKEL}, where we display $e_{\nu}$ values up to $^{78}$Ni. We find that the isotopic trend is inverted for $^{78}$Ni, whose $e_\nu$ are in most cases bigger than those for $^{68}$Ni. This is less the case for the $e_{\nu}$ computed in FTDA, as seen from the values in parentheses in the last column of Table~\ref{tab:Nickel_Neutron_RPA}, and also for the $e_{\pi}$ values in the last column of Table~\ref{tab:Nickel_Proton_RPA}. However, the enhancement is systematic and indeed we have checked for both neutron and proton effective charges that  is not related to the divergences in the energy denominators of Eq.~(\ref{eq:t_expansion}). The only exception is given by the $\nu f_{\frac{7}{2}}$ orbital, whose energy denominator is responsible for the drastic change in the $e_{\nu \! f_{\!\frac{7}{2}} \, \! \nu \! p_{\!\frac{3}{2}}}$ when moving from FTDA to FRPA, with the former having an unphysical (negative) value.  The difference between FRPA and FTDA effective charges in $^{78}$Ni could be then an hint of poor convergence of the calculation for that isotope at $N_{\text{max}}$=~11.

\begin{figure}[t]
  \centering
    \subfloat{\includegraphics[scale=0.33]{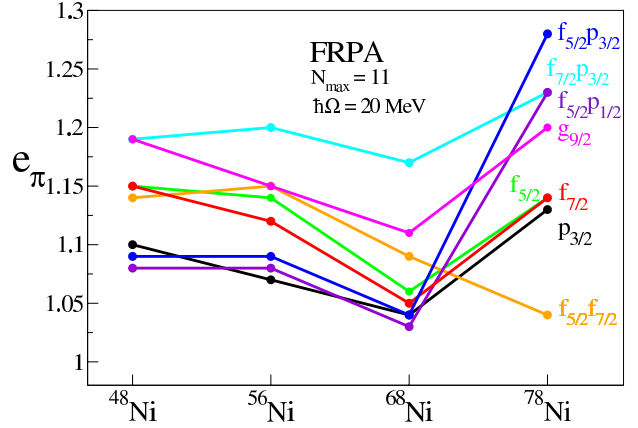}}
  \caption{ Values of Nickel $e_\pi$ for selected orbitals, displayed in Table~\ref{tab:Nickel_Proton_RPA}. Straight lines are a guide to  the eye.}
  \label{e_p_plot_NICKEL}
\end{figure}

 We consider proton orbitals for all four isotopes in Table~\ref{tab:Nickel_Proton_RPA}, and found effective charges 
$e_\pi \simeq $ 1.1-1.2, except for $p$-shell orbitals in $^{68}$Ni, having $e_\pi \simeq $ 1.0. The possible lack of convergence for the $^{78}$Ni seems to be confirmed also for the $e_\pi$ values, which tend to  diverge systematically from the smooth trend of the three lighter Nickel isotopes. For completeness, we plot in Fig.~\ref{e_p_plot_NICKEL} the proton effective charges for selected orbitals, including  the  $^{78}$Ni values showing a departure from the expected isotopic trend.

\section{\label{concl}Conclusions}
We have calculated E2 effective charges for Oxygen and Nickel isotopes, respectively for the  $0p\, 1s\, 0d$ and $1p\,0f\,0g_{\frac{9}{2}}$ valence spaces.  The calculations were based on the saturating realistic interaction NNLO$_{\text{sat}}$. The very large harmonic oscillator space allows to compute renormalization corrections to the bare charges by taking into account the coupling of single nucleons to collective excitations beyond the assumed shell-model valence space. The core-polarization effects induced by the electromagnetic field, have been described in an \emph{ab initio} fashion according to the SCGF formalism. This method allows a systematic improvement of the description of the virtual phonon excitations, within a many-body formalism capable to resum non-perturbatively both particle-particle, hole-hole and particle-hole correlations with the inclusion of ladder and ring diagrams, respectively.
The approach allows to choose any specific valence space and  to block correlations effects from the ISCs belonging to the valence space itself, so that they would not be double counted in successive shell-model calculations.
The formalism  can be generally applied to any one-body external field operator that transfers a generic angular momentum~$\lambda$.

For each of the isotopes considered, we have found that effective charges depend significantly on the orbits with variations that are more pronounced for neutrons than protons. Moreover, for both chains of isotopes, the values of $e_{\nu}$ for fixed orbits are isospin dependent, displaying a decreasing trend when increasing the number of neutrons. In this respect, our calculations confirm the effect of polarization quenching for nucleons when the Fermi energy is approaching  the neutron drip line.
In general, the E2 effective charges for the four Oxygen nuclei (see Tables~\ref{tab:Oxy_Neutron} and~\ref{tab:Oxy_Proton}) are compatible with the phenomenological values adopted for shell-model calculations of $0p\, 1s\, 0d$ isotopes. 
For the  Nickel isotopes (see Tables~\ref{tab:Nickel_Neutron_RPA} and~\ref{tab:Nickel_Proton_RPA}), $e_{\pi}$ polarization corrections are found that are smaller than the standard values adopted ones for $1p\,0f\,(0g_{\frac{9}{2}})$ nuclei. This finding is also in line with the recent literature for isotopic chains in the Nickel 
region~\cite{Brown1974,Hoischen2011,Allmond2014,Loelius2016}.
The results for  $^{78}$Ni need to be confirmed by further calculations with a better assessment of the convergence in terms of the size of the model space, beyond the  $N_{\text{max}}$= 11 adopted here.  Forthcoming  advances in computational methods and resources may allow a better study of this case in the near 
future~\cite{Soma2018}.

The present work is based on a fully \textit{ab initio} treatment of correlations done at the ADC(3)/FTDA and FRPA level and it treats the external field $\hat\phi^{(\lambda)}$  by using linear response theory--as usually done--which is well justified by the smallness of the fine-structure constant.
The nuclear structure effects are instead computed in a fully non-perturbative way. Nonetheless, the formalism for the computation of shell-model effective charges and interactions is still at an early stage and further developments will be required to exploit the capabilities of this approach. In particular, one important open question is the most appropriate choice for the orbits representing the valence space and how the microscopic correlations effects should be mapped into it.  In this work, we have chosen orbits from an optimised reference state that best describes quasiparticle states near the Fermi surface while maintaining an orthonormal basis. While this is an ideal choice from the physics point of view, the present formulation can lead to diverging denominators and consequently to an uncontrolled behaviour of some calculated effective charges.  Even with these limitations, it was already possible to extract the general trends of the shell-model effective charges as a function of the isospin asymmetry.  Our results substantially confirm the empirical findings 
of Refs.~\cite{Sagawa2004}-\cite{Loelius2016} for nuclei in the same regions of the nuclear chart but from a microscopic, \textit{ab initio} inspired, point of view.

This work is to be considered as a first step toward developing a consistent formalism for mapping  \textit{ab initio} SCGF theory into a shell-model framework.
While the formalism  needs to be evolved to resolve the above divergence issues and to include more sophisticated many-body truncations, our results clearly show that the approach is viable. Such extensions will be the object of future work.


\begin{acknowledgments}
We thank Petr Navr\'{a}til for providing the matrix elements of the NNLO$_{\text{sat}}$ Hamiltonian. 
C.B. thanks T. Otsuka for inspiring this work. This research is supported by the United Kingdom Science and Technology Facilities Council (STFC)
under Grants No. ST/P005314/1 and No. ST/L005816/1
 Calculations were performed performed using the DiRAC Data Intensive service at Leicester (funded by the UK BEIS via STFC capital grants ST/K000373/1 and ST/R002363/1 and STFC DiRAC Operations grant ST/R001014/1).
\end{acknowledgments}





\bibliography{EFF_CH}

\end{document}